# Rydberg Excitons and Trions in Monolayer MoTe$_2$

Souvik Biswas[1,11], Aurélie Champagne[2,3], Jonah B. Haber[3], Supavit Pokawanvit[4,5], Joeson Wong[1,11], Hamidreza Akbari[1], Sergiy Krylyuk[6], Kenji Watanabe[7], Takashi Taniguchi[8], Albert V. Davydov[6], Zakaria Y. Al Balushi[1], Diana Y. Qiu[9], Felipe H. da Jornada[5], Jeffrey B. Neaton[2,3,10], Harry A. Atwater[1,11]

Affiliations

1. Thomas J. Watson Laboratory of Applied Physics, California Institute of Technology, Pasadena, CA 91125, USA
2. Materials and Chemical Science Division, Lawrence Berkeley National Laboratory, Berkeley, California 94720, USA
3. Department of Physics, University of California Berkeley, Berkeley, California 94720, USA
4. Department of Applied Physics, Stanford University, Stanford, CA 94305, USA
5. Department of Materials Science and Engineering, Stanford University, Stanford, California 94305, USA.
6. Materials Science and Engineering Division, National Institute of Standards and Technology, Gaithersburg, Maryland 20899, USA
7. Research Center for Functional Materials, National Institute for Materials Science, 1-1Namiki, Tsukuba 305-0044, Japan
8. International Center for Materials, Nanoarchitectonics, National Institute for Materials Science, 1-1 Namiki, Tsukuba 305-0044, Japan
9. Department of Mechanical Engineering and Materials Science, Yale University, New Haven, Connecticut 06520, United States
10. Kavli Energy Nanosciences Institute at Berkeley, Berkeley, CA 94720, USA
11. Kavli Nanoscience Institute Pasadena, CA 91125, USA

**Abstract**

Monolayer transition metal dichalcogenide (TMDC) semiconductors exhibit strong excitonic optical resonances which serve as a microscopic, non-invasive probe into their fundamental properties. Like the hydrogen atom, such excitons can exhibit an entire Rydberg series of resonances. Excitons have been extensively studied in most TMDCs (MoS$_2$, MoSe$_2$, WS$_2$ and WSe$_2$), but detailed exploration of excitonic phenomena has been lacking in the important TMDC material molybdenum ditelluride (MoTe$_2$). Here, we report an experimental investigation of excitonic luminescence properties of monolayer MoTe$_2$ to understand the excitonic Rydberg series, up to 3s. We report significant modification of emission energies with temperature (4K to 300K), quantifying the exciton-phonon coupling. Furthermore, we observe a strongly gate-tunable exciton-trion interplay for all the Rydberg states governed mainly by free-carrier screening, Pauli blocking, and band-gap renormalization in agreement with the results of first-principles GW plus Bethe-Salpeter equation approach calculations. Our results help bring monolayer MoTe$_2$ closer to its potential applications in near-infrared optoelectronics and photonic devices.

**Introduction**

Excitons[1], excitations which consist of bound electron-hole pairs, in monolayer transition-metal dichalcogenide (TMDC) semiconductors are a suitable platform to investigate a rich variety of condensed-matter phenomena – such as Mott insulators[2,3], Wigner crystals[4], and light-induced magnetic phases[5] – via optical spectroscopy due to their high binding energy and large oscillator strength[6–10]. The pronounced optical resonances due to excitons in TMDCs lie below the electronic gap and arise from their atomically thin, two-dimensional nature, which features pronounced quantum confinement and weak dielectric screening[1,7,9–11]. In monolayer TMDCs, a valley degree of freedom[12–14] emerges from the crystal structure with $C_3$ and broken inversion symmetries, which give rise to lowest-energy excitonic states with two-fold degeneracy and displaying opposite chiral selection rules. Beyond the lowest-energy optical excitations, excitons can also exist in hydrogenic internal excited states[15–17], known as Rydberg excitons, which by virtue of their relatively larger wavefunction[7,18] offer a sensitive probe of exciton-electron, exciton-exciton, and other quasiparticle interactions, making them attractive candidates for optical quantum sensing[19–21]. Additionally, Rydberg excitons offer a way to realize giant light-matter interactions (similar to Rydberg atoms) and can be studied in cavity quantum electrodynamics and nonlinear optical measurements[22,23]. While typically observed in resonant reflection measurements[11,24], a number of recent studies on extremely high-quality TMDC samples have reported signatures of Rydberg excitons in photoluminescence (PL)[16,25–28]. Although such states have been extensively characterized in monolayer $MoS_2$, $MoSe_2$, $WS_2$ and $WSe_2$, [15,16,24–30] they have not been well explored in monolayer $MoTe_2$ [30].

The 2H phase of monolayer $MoTe_2$ is semiconducting, with the smallest bandgap (in the near infrared) among the Mo-based TMDC materials[31–38]. Several studies have reported a phase transition to the metallic-1T' phase, under high carrier doping conditions, which may be useful in phase change photonics[39,40]. The optical properties of $MoTe_2$ change dramatically under the extreme conditions of carrier doping, but even at lower carrier doping (~$10^{11}$ cm$^{-2}$), the excitonic properties are significantly altered and yield information about quasiparticle and exciton interactions and define exciton-electron dynamics[26–28,41]. Because of the presence of heavy tellurium atoms, the spin-orbit coupling effects (~230 meV for valence band and ~43 meV for conduction band) [A. Champagne et al., submitted] and bright-dark A-exciton splitting (~25 meV)[35] in $MoTe_2$ are much more significant compared to the other Mo-based TMDCs. Additionally, $MoTe_2$ is one of the few van der Waals materials to emit near the silicon band-edge and hence, an accurate understanding of the photo-physics of the entire Rydberg series under different conditions of excitation density, temperature and doping can guide future development of near-infrared optoelectronic and photovoltaic components – such as detectors, modulators, and light emitting diodes.

In this work, we report results of the experimental characterization of the optical properties of electrostatically gated monolayer $MoTe_2$, probed via photoluminescence measurements. Combining high-quality heterostructures and a resonant back-reflector geometry, we identify different optical transitions corresponding to the excitonic Rydberg series. The evolution of emission as a function of temperature, reveals a semiconductor-like behavior with quantitative estimation of zero-temperature energies and Rydberg exciton-phonon coupling strengths. By controlling the charge density in the monolayer $MoTe_2$ from charge neutrality up to electron/hole densities of ~$10^{12}$ cm$^{-2}$ we find strong modulation of optical transitions and continuous tuning of the ground and excited state excitonic manifold – which is computed and illustrated in Fig. 1(a). We also perform first-principles calculations based on many-body perturbation theory (MBPT) to obtain the excited-state properties of monolayer $MoTe_2$ including many-electron interactions. First principles GW plus Bethe Salpeter equation (GW-BSE) calculations with a new plasmon pole model developed [A. Champagne et al., submitted] to account for the dynamical screening of carriers show that the strong tunability is attributed to enhanced screening of the excitonic states from the increased electron density as well as phase space filling which leads to Pauli blocking of optical transitions. Our *ab initio* calculations also capture well the trion binding energy close to charge neutrality. Additionally, a linear linewidth broadening is observed which is attributed to enhanced exciton-electron scattering with increasing carrier density, in qualitative agreement with explicit calculations that consider the scattering of excitons to the degenerate Fermi sea.

**Results**

To efficiently probe the optical properties of monolayer MoTe$_2$ samples, we adopt a Salisbury-screen[42] geometry (details are in the SI – (B)), shown schematically in Fig. 1(b). The MoTe$_2$ is placed approximately a quarter wavelength away from a back-reflector of optically thick gold which also acts as the bottom electrode, to cause a destructive interference of the electromagnetic field at the monolayer, thereby enhancing light-matter interaction. This configuration results in a mild Purcell enhancement of the emission[43] ($F_p \sim 2$, where $F_p$ is the Purcell factor, SI (C)) and allows efficient tuning of the Fermi level in MoTe$_2$ with the application of a gate voltage across the bottom hBN.

Measurements of the spatial dependence of PL at T = 4K yield bright, uniform emission from the monolayer regions of the device. Fig. 1(c) shows integrated PL counts, $I_{PL} = \int_{\lambda=800nm}^{1150nm} I(\lambda)d\lambda$, over a bandwidth from 800nm (~1.55eV) to 1150nm (~1.07eV). The bilayer regions exhibit lower emission and a broader peak, while very faint emission is seen from the multilayer regions. There have been reports investigating whether bilayer can become direct gap semiconductor at lower temperatures using emission spectroscopy because of the similar photoluminescence quantum yield[36], $PLQY = \frac{\gamma_r}{\gamma_r + \gamma_{nr}}$, where $\gamma_r, \gamma_{nr}$ are the radiative and non-radiative rates, respectively. From our measurements (shown in SI (D)), we find the ratio of the PLQY of the monolayer to bilayer to be three, indicating that the indirect to direct gap transition might happen when MoTe$_2$ is thinned to a bilayer. A representative PL spectrum (Fig. 1(d)) from one of the brightest monolayer spots shows sharp emission around 1.172 eV and 1.149 eV with linewidths of 7.13 and 8.77 meV, respectively, which we attribute to the A1s exciton and trion, as reported previously[31,34–38]. Some of the cleanest regions of the sample show extremely narrow linewidths, the narrowest obtained being ~4.48 meV, indicating high sample quality (shown in SI (E)).

Ab initio GW-BSE calculations using a modified plasmon pole model to account for dynamical screening associated with free carriers [A. Champagne et al., submitted] support our understanding of the experimental spectra and enable prediction of important optical properties such as the quasiparticle band gap, optical resonance energies, and exciton binding energies. Within this formalism, one- and two-particle excitations can be calculated using state-of-the-art GW and GW-BSE approach, respectively. Computational details are reported in SI ((O)-(S)) and elsewhere [A. Champagne et al., submitted]. In monolayer MoTe$_2$, the lowest interband excitonic transition is bright and occurs at the degenerate K and K' points in the Brillouin zone (Fig. 1(a)). Due to spin-orbit coupling effects, a splitting of both the valence band maximum and conduction band minimum occurs, resulting in two distinct series of excitons, typically labeled as A and B excitons. The optical gap computed with the first-principles GW-BSE approach, and which corresponds to the lowest-energy A1s exciton energy, is found at $E_{opt} = 1.09$ eV, in good agreement with the experimentally measured gap. The computed GW quasiparticle gap is $E_g = 1.58$ eV (see SI (Q) for details). Thus, we predict an exciton binding energy of $E_b = E_g - E_{opt} = 490$ meV for the lowest energy optically-bright excited state. Because of strong Coulomb interactions in low dimensions, charged excitons (trions) are expected to form in monolayer MoTe$_2$. We calculate the binding energy of the negatively charged exciton from first principles by solving the corresponding equation of motion for three-quasiparticle correlated bound states (see SI (S) for details) and obtain 20.6 meV (see SI (S)), in agreement with former reports[31] and measured values of ~23 meV.

Additional luminescence peaks are seen at higher energies at 1.269 eV, 1.29 eV and 1.315 eV, and have been identified as the A2s trion, A2s exciton and A3s exciton, respectively, in accordance with previous reports on MoTe$_2$[30] and other TMDCs. The assignment of these excitonic and trionic peaks is in accordance with nomenclature which reflects how the exciton wavefunction transforms under the crystal symmetry in analogy with the hydrogen atom[44]. Our ab-initio GW-BSE calculations (see SI (O), (Q) for details) predict higher excited state excitons at 1.31 eV, 1.35 eV, and 1.43 eV, corresponding to the A2s, B1s, and A3s excitons, respectively. The slight discrepancy in the peak energies and re-ordering of the B1s and A3s exciton is likely related to the enhanced screening of MoTe$_2$ by the hBN dielectric, which is not considered in the calculations. The observation of Rydberg states up to 3s enables investigation of stronger Rydberg interactions, as well as photon-matter coupling, in MoTe$_2$. The linewidths observed for these excited states are exceptionally narrow, ~10 meV and ~25 meV for the 2s and 3s exciton, respectively. Additionally, the brightness of the 2s state is ~10% of the 1s state which is comparable to or higher than that for other TMDCs.

To verify that the emission is excitonic in nature, we performed pump-power dependent photoluminescence measurements. We scanned the incident pump fluence over 2 decades in intensity and observed an increase in the emission intensity (Fig. 2(a)). A mild spectral broadening is associated with increasing pump density, originating from enhanced exciton-exciton interactions. We analyze the peaks by fitting to a Lorentzian lineshape profile, $I_{PL} = \sum_{i=Rydberg\ states} \frac{\frac{A_i \Gamma_i}{2}}{(\omega-\omega_i)^2 + \left(\frac{\Gamma_i}{2}\right)^2}$ – (E1), where $A_i$ is the oscillator strength, $\Gamma_i$ is the broadening (full-width at half maximum) and $\omega_i$ is the resonance frequency of each resonance, respectively. We can extract the integrated PL intensity, $\int I_{PL,i}(P)\ d\omega = C_{0,i} P^{\alpha_i}$, where $P_i$ is the incident power and $\alpha_i$ is the exponent for each resonance, $I_{PL,i}$ is given by equation E1 and $C_{0,i}$ is a dimensionless constant as a function of pump power. Near linear scaling is seen for all the excitonic states as shown in Fig. 2(b), plotted in semi-log scale, with exponents as $\alpha = 1, 0.93, 0.92$ for 1s, 2s and 3s states, respectively. This excludes any defect related emission as no saturation or non-linearity is observed over 2 orders of magnitude of incident pump power.

Rydberg excitons in MoTe$_2$ also show strong temperature dependence, consistent with previous observations in other TMDCs. While the lowest energy state has been investigated for MoTe$_2$, there is lack of knowledge about the excited state dynamics with temperature. Our measurements, in Fig 2(c) and (d) for the 1s and 2s exciton, respectively, show a redshift for excitonic states with increasing temperature which can be modeled with a semi-empirical semiconductor bandgap dependence of the form $E_{exc}(T) = E_{exc}(0) - S\langle \hbar \omega \rangle \left[\coth\left(\frac{\langle \hbar \omega \rangle}{k_B T}\right) - 1\right]$, where $E_{exc}(0)$ is the resonance energy at zero temperature limit, $S$ is a dimensional constant, $k_B$ is the Boltzmann constant and $\langle \hbar \omega \rangle$ is the average phonon energy[34]. From the fits, we extract the parameters summarized in Table T1, also shown in Fig. 2(e), (f). Furthermore, the PLQY drops with increasing temperature, which is attributed to an increase in accessible non-radiative decay channels from the phonon contributions (evident from the linewidth broadening with increasing temperature, shown in SI (I)), while the radiative contribution remains constant. The zero-limit exciton energy indicates the Rydberg state energy levels, in close agreement with ab initio GW-BSE computed energy levels (A1s: 1.09 eV and A2s: 1.31 eV), which correspond to $T = 0K$. Interestingly, the relative intensity of the 2s exciton state with respect to the 1s state grows with increasing temperature (see SI (I)), possibly stemming from weaker coupling with the phonons.

Table T1. Experimentally measured zero-temperature exciton energy and exciton-phonon coupling parameters for the Rydberg states.

| State index | Energy ($E_{exc}(0)$) (eV) | S | $\langle \hbar \omega \rangle$ (meV) |
|---|---|---|---|
| 1s | 1.176 ± 0.019 | 1.22 ± 0.434 | 8.7 ± 7.009 |
| 2s | 1.292 ± 0.018 | 1.08 ± 0.304 | 6.7 ± 5.825 |

Reduced dielectric screening and strong electron-hole Coulomb interactions in two dimensional semiconductors make their electronic and optical properties highly sensitive to their dielectric environments[9,11,17,41,45–47]. In particular, the presence of free carriers can significantly affect the electronic landscape of a TMD monolayer[26–29,31,41]. A key finding of our study, summarized in Fig. 3, is the carrier dependence of the exciton-electron interaction, as quantified by the gate voltage dependence of the exciton and trion emission properties. We first focus near the 1s exciton resonance illustrated in Fig. 3(a)-(c). A false color map shows the evolution of the 1s exciton and trion peaks as a function of applied gate voltages ($V_g = -10V$ to $10V$). At very low voltages the neutral exciton peak dominates in emission, but with a very small change in the carrier density the trion peak rapidly emerges as a dominant feature on either side of $V_g \sim -0.65V$ (which is identified as the charge neutral condition from the peak in neutral 1s exciton emission intensity). At higher voltages the emission from the neutral exciton is completely suppressed. While trion emission grows in intensity for higher voltages, it eventually saturates and shows a slight reduction at even higher voltages. Such changes are better visualized in the derivative of PL with respect to energy ($\frac{dPL}{dE}$), shown in Fig. 3(b) and (e) for different regions of the gate voltage. The resonances corresponding to the 2s and 3s excitonic states show a qualitatively similar doping dependence (Fig. 3(d)-(f)). Line-cuts corresponding to near charge neutral condition and finite doping showing strong exciton and trion emission spectrum are plotted in Fig. 3(c) and (f) for the 1s and 2s, 3s states, respectively.

To quantitatively understand the doping induced changes in the emission dynamics, the spectrum is fitted to a sum of multiple Lorentzian features, as given by (E1), corresponding to the different exciton and trion states. The evolution of the peak intensity, linewidth and energy are then extracted as a function of carrier density, with the results presented in Fig. 4. Fig. 4(a) and (b) quantify the changes in the PL intensity of the different exciton and trion states, as discussed previously. A crossover-density ($N_c$) is defined where the exciton and trion intensities overlap and is identified to be $V_g = 0.296V$ and $V_g = -1.36V$ on the electron and hole side, respectively, for the 1s state. This corresponds to (from a parallel plate capacitor model – see SI (G)) charge densities of $N_c^- = 2.08 \times 10^{11}$ cm$^{-2}$ and $N_c^+ = 1.54 \times 10^{11}$ cm$^{-2}$, respectively. Additionally, as seen in Fig. 4(e), the exciton slightly blue shifts (2s much more than 1s) with increasing charge density, while the trion redshifts (see SI (M)). A qualitatively similar trend is seen for the features corresponding to the 2s exciton state and a crossover-density of $V_g = 1.24V$ and $V_g = -2.42V$, corresponding to $N_c^- = 4.19 \times 10^{11}$ cm$^{-2}$ and $N_c^+ = 3.95 \times 10^{11}$ cm$^{-2}$ is identified on the electron and hole side, respectively. The emission strength from the 3s trion state, which appears red-shifted to the 3s excitonic state, is not high enough to perform further quantitative analysis. However, from the derivative of PL measurements in Fig. 3(e) and (f), it is clear that a similar qualitative picture also holds true for the 3s state. Further studies with magnetic fields are required to study quantitative dynamics of the even higher states, so that the visibility is improved. Our observations are consistent with previous reports of gate-tunable exciton and trion intensities in other TMDCs[26–28].

The energy differences ($\Delta E = E_{exciton} - E_{trion}$) between the exciton and trion exhibit an unusual gate dependence and show striking difference between the 1s and 2s states (Fig 4(c)). An overall slight blue shift of the energy difference $\Delta E_{1s} \sim 1$ meV, over a doping density of $n \sim 1 \times 10^{12}$ cm$^{-2}$ is seen. A linear fit for $\Delta E_{1s}$ reveals a rate of change in the exciton energy of $0.103 \frac{\text{meV}}{10^{11} \text{ cm}^{-2}}$ ($0.118 \frac{\text{meV}}{10^{11} \text{ cm}^{-2}}$) for the electron (hole) doping. A zero-density limit of the energy shift provides the trion binding energy which is 21.94 meV (22.14 meV) for the negative (positive) trion. The energy shift is highly exaggerated for the 2s state, where a much larger shift of $\Delta E_{2s} \sim 10$ meV is seen over a smaller doping density of

n~5x10$^{11}$ cm$^{-2}$. A similar analysis yields 1.57 $\frac{meV}{10^{11} \text{ cm}^{-2}}$ (2.56 $\frac{meV}{10^{11} \text{ cm}^{-2}}$) for the electron (hole) doped case. The trion binding energies are estimated to 18.14 meV (13.76 meV) for the negative (positive) side. The lower binding energy of the 2s trion compared to the 1s state follows from the lower binding energy of the corresponding neutral state. This unusually larger energy shift is understood as stemming from the larger wavefunction of the 2s state and thus, higher susceptibility to the electronic landscape which also is evident from the stronger dependence of the oscillator strength with doping density of the 2s state as compared to the 1s state. These results are illustrated in Fig. 4(c) and summarized in Table T2. In general, we expect the sensitivity of self-doping to increase dramatically with even higher lying Rydberg states.

Table T2. Experimentally measured binding energy and energy shifts for Rydberg trions.

| State index | $E_{b,trion^+}$ (meV) | $E_{b,trion^-}$ (meV) | $\frac{\Delta E}{\Delta n^+}$ (meV/10$^{11}$ cm$^{-2}$) | $\frac{\Delta E}{\Delta n^-}$ (meV/10$^{11}$ cm$^{-2}$) |
|---|---|---|---|---|
| 1s | 22.14 ± 0.20 | 21.94 ± 0.10 | 0.118 ± 0.03 | 0.103 ± 0.02 |
| 2s | 13.67 ± 0.12 | 18.14 ± 0.09 | 2.56 ± 0.16 | 1.57 ± 0.19 |

We also measure the linewidth evolution (Fig. 4(d)) and observe that neutral exciton states exhibit linewidth broadening as a function of doping density. Additionally, while the 2s trion broadens, the 1s trion remains nearly unchanged with increasing carrier concentration.

**Discussion.**

To better understand doping induced changes in the optical properties we use first-principles GW and GW-BSE calculations with a modified plasmon pole model [A. Champagne et al., submitted] to compute the exciton spectrum under different carrier densities. For optical excitations close to the band edge, we find two main effects to support our experimental observation: (i) a doping-independent ground exciton energy (Fig.5(b)) and (ii) a suppression of the exciton oscillator strength (Fig.5(c)). Numerical results are reported in Table T3.

Table T3 – Computed Rydberg exciton binding energy and relative dipole moment as a function of doping density.

| Doping density (cm$^{-2}$) | A1s | | A2s | | A3s | |
|---|---|---|---|---|---|---|
| | $\Delta E_b$ (meV) | Rel. Dipole Moment | $\Delta E_b$ (meV) | Rel. Dipole Moment | $\Delta E_b$ (meV) | Rel. Dipole Moment |
| 0 | 0 | 1 | 0 | 0.25 | 0 | 0.09 |
| 2.3x10$^{11}$ | -116 | 0.58 | -78 | 0.08 | -90 | 0.03 |
| 1.6x10$^{12}$ | -246 | 0.28 | -173 | 0.03 | -192 | 0.02 |
| 3.0x10$^{12}$ | -301 | 0.10 | Not detectable | | | |
| 4.5x10$^{12}$ | -353 | 0.05 | | | | |
| 5.9x10$^{12}$ | -366 | 0.02 | | | | |
| 8.7x10$^{12}$ | -368 | 0.01 | | | | |

The evolution of the exciton energy with increasing doping density arises from an interplay of various effects[48–50]. In the low-doping regime, the doping-independent exciton energy results from a compensation between the band gap renormalization and exciton binding energy reduction (Fig.5(a) and (b)), expected from the reduced electron-hole Coulomb interaction. At higher doping concentration, the ground exciton peak is expected to blueshift slightly, as the exciton binding energy saturates, while the quasiparticle gap slightly increases due to an increase of the energy continuum with the free carrier concentration[49]. In addition, as the doping density increases, the exciton delocalizes in real space (exciton wave function reported in SI (R)), and Pauli blocking prevents transitions around the K valley, which is eventually reflected in a decrease of the oscillator strength of the exciton peak, as shown in Fig.5(c). This argument can be related to the intrinsically lower oscillator strength (and by reciprocity, lower PLQY) of the higher order Rydberg states – also due to a more delocalized wave-function in real space. Similarly, the exciton binding energy and oscillator strength of the A2s and A3s excited states decrease rapidly, and the corresponding peaks quickly vanish above a doping density of $2 \times 10^{12}$ cm$^{-2}$.

The average lifetime, $\tau$, of an unstable particle is related to the decay rate $\gamma$, as $\tau = \frac{1}{\gamma}$. The PL linewidth $L$, obtained as $L = \frac{\hbar}{\tau} = \hbar \gamma$, provides information about intrinsic contributions from radiative exciton lifetime and dephasing from exciton-phonon scattering, as well as extrinsic inhomogeneous broadening effects (e.g., doping, defects, substrate-induced disorder). In MoTe$_2$ monolayer, the bright exciton state is energetically below the dark states, which, at low temperature, prevents scattering towards intervalley exciton states that would require the absorption of a phonon. Therefore, at low temperature, the intrinsic contributions to the linewidth are dominated by radiative exciton decay[51]. Using Fermi's golden rule[52], we compute a radiative exciton lifetime of 0.3 ps, corresponding to a radiative linewidth of 2.2 meV for the ground exciton. The discrepancy with the experimental zero-doping linewidth of 7.13 meV comes from inhomogeneous broadening effects, such as the presence of defects or substrate effects. Furthermore, due to the presence of a back reflector which gives rise to a slight Purcell enhancement in emission, it is expected that the enhanced radiative rate is higher than the computed one in vacuum (by approximately ~2). With increasing doping density, charged excited states, known as trions, emerge and couple to the excitons. Using the microscopic many-body theory developed in previous studies[50], we expect an approximately linear exciton linewidth broadening with doping density (~8.7 meVcm$^{-2}$, in our measurements) due to enhanced exciton-electron scattering.

**Conclusions**

In summary, we report on the optical luminescence features of a monolayer MoTe$_2$, including Rydberg excitons up to 3s states, by combining results of experimental photoluminescence measurements and first principles calculations. We observe a linear dependence of the exciton peak energy with incident pump fluence and a red shift with increasing temperature, following a semi-empirical semiconductor relationship. The optical response can further be modulated with gate voltage, with an efficient exciton to trion conversion. With increasing doping density, we predict (i) a reduction in the exciton oscillator strength and (ii) a near-constant (mild blueshift) exciton energy, supporting our experimental measurements. Our understanding of MoTe$_2$ photo-physics creates a foundation for understanding and design of future optoelectronic devices in the near infrared.

**Methods**

**Fabrication** - SiO$_2$ (285 nm)/Si chips were cleaned with ultrasonication in acetone and isopropanol for 30 min. each, followed by oxygen plasma treatment at 70 W, 300 mTorr for 5 min. Monolayer MoTe$_2$, few layer hBN and graphene flakes were directly exfoliated using Scotch-tape at 100 ºC to increase the yield of monolayer flakes. Monolayer thickness was initially identified using optical contrast (~7 % contrast per layer) and later verified with atomic force microscopy. hBN thickness was confirmed with atomic force microscopy. Flakes were assembled with a polycarbonate/polydimethylsiloxane (PC/PDMS) stamp with pick-up at temperatures between 80 ºC -110 ºC. The entire heterostructure stack was dropped on prefabricated gates at 180 ºC. The polymer was washed off in chloroform overnight, followed by isopropanol for 10 min. Given the air-sensitive nature of MoTe$_2$ monolayers, exfoliation, identification of MoTe$_2$ flakes and stacking of the entire heterostructures were done in a nitrogen purged glovebox with oxygen and moisture levels below 0.5 ppm. Gates were fabricated on SiO$_2$ (285 nm)/Si chips with electron beam lithography (PMMA 950 A4 spun at 3500 rpm for 1 min. and baked at 180 $^{\circ}$C, 10 nA beam current and dosage of 1350 $\mu$C/cm$^2$ at 100 kV), developing in methyl isobutyl ketone: isopropanol (1:3) for 1 min., followed by isopropanol for 30 s and electron beam evaporating 5 nm Ti/95 nm Au at 0.5 Å/$s$ deposition rate. Liftoff was done in warm acetone (60$^{\circ}$C) for 10 min., followed by rinse in isopropanol for 5 min. The gates were precleaned before drop-down of heterostructure by annealing in high vacuum (2x10$^{-7}$ Torr) at 300 $^{\circ}$C for 6 h. The chip was then wire-bonded with Aluminum wires on to a custom home-made printed circuit board.

**Optical spectroscopy** - Low temperature confocal photoluminescence measurements were performed in an attoDRY800 closed-cycle cryostat at base pressures of <2x10$^{-5}$ mbar. Sample was mounted on a thermally conducting stage with Apiezon glue and stage was cooled using a closed-cycle circulating liquid helium loop. Temperature was varied between ~4K and 300K. A 532 nm (Cobolt) continuous wave laser was used as the excitation source with power ranging between 10 nW and 1 mW, focused to a diffraction limited spot. Emission was collected in a confocal fashion with a cryogenic compatible apochromatic objective with an NA of 0.82 (for the visible and NIR range, LT APO VISIR) and dispersed onto a grating-based spectrometer (with 150 grooves per mm with a Silicon CCD) – Princeton Instruments HRS 300. Voltage was applied using a Keithley 2400. Data was acquired with home-written MATLAB codes.

**Crystal growth** - MoTe$_2$ single-crystals were grown by the chlorine-assisted chemical vapor transport (CVT) method. A vacuum-sealed quartz ampoule with polycrystalline MoTe$_2$ powder and a small amount of TeCl$_4$ transport agent (4 mg per cm$^3$ of ampoules' volume) was placed in a furnace containing a temperature gradient so that the MoTe$_2$ charge was kept at 825 °C, and the temperature at the opposite end of the ampoule was about 710 °C. The ampoule was slowly cooled after 6 days of growth. The 2H phase of the obtained MoTe$_2$ flakes was confirmed by powder X-ray diffraction and transmission electron microscopy studies. Flakes obtained from the aforementioned source as well as commercially available MoTe$_2$ (2D Semiconductors) were investigated with similar results.

**AFM imaging** - Atomic Force Microscopy was performed using Bruker Dimension Icon in tapping mode. Data analysis was performed in MATLAB.

**Acknowledgements**

The experimental measurements were obtained under support from the US Department of Energy Physical Behavior of Materials program, under grant DE-FG02-07ER46405. The electronic structure calculations for the development of the plasmon-pole model is supported by the US Department of Energy Center for Computational Study of Excited-state Phenomena in Energy Materials (C2SEPEM) - (Many-body perturbation theory calculations) and the Theory of Materials FWP (development of the plasmon-pole model), under contracts No. DE-AC02-05CH11231. Computational resources are provided by the National Energy Research Scientific Computing Center (NERSC) and the Texas Advanced Computing Center (TACC) at The University of Texas at Austin, funded by the National Science Foundation (NSF) award 1818253, through allocation DMR21077. A.C. acknowledges the support from Wallonie Bruxelles International. A.V.D. and S.K. acknowledge support through the Material Genome Initiative funding allocated to the National Institute of Standards and Technology. K.W. and T.T. acknowledge support from the Elemental Strategy Initiative conducted by Ministry of Education, Culture, Sports, Science and Technology (MEXT grant JPMXP0112101001), Japan Society for the Promotion of Science (JSPS KAKENHI grant JP20H00354), and Centers of Research Excellence in Science and Technology (CREST grant JPMJCR15F3), Japan Science and Technology Agency (JST). Authors thank Eoin Caffrey and Dr. Pin Chieh Wu for their support. D.Y.Q. acknowledges support by a 2021 Packard Fellowship for Science and Engineering from the David and Lucile Packard Foundation.

**Author contributions**

S.B. and H.A.A conceived the project. S.B. fabricated $MoTe_2$ gated heterostructures, performed characterization and optical measurements and analysis of the data. J.W., H.A. and Z.Y.A.B. assisted in optical measurements and discussions. A.C. led the MBPT calculations with help from J.B.H. and S.P. and inputs from F.H.J., D.Y.Q. and J.B.N. S.K. and A.V.D. provided $MoTe_2$ crystals and K.W. and T.T provided hBN crystals. S.B. wrote the manuscript with input from all authors. H.A.A. supervised the project.

**ORCID ID:**

Souvik Biswas - 0000-0002-8021-7271
Aurélie Champagne - 0000-0002-6013-2887
Jonah B. Haber - 0000-0001-8192-9994
Supavit Pokawanvit - 0000-0002-7799-9452
Joeson Wong - 0000-0002-6304-7602
Hamidreza Akbari - 0000-0002-6073-3885
Sergiy Krylyuk - 0000-0003-4573-9151
Kenji Watanabe - 0000-0003-3701-8119
Takashi Taniguchi - 0000-0002-1467-3105
Albert V. Davydov - 0000-0003-4512-2311
Zakaria Y. Al Balushi - 0000-0003-0589-1618
Felipe H. da Jornada - 0000-0001-6712-7151
Diana Y. Qiu - 0000-0003-3067-6987
Jeffrey B. Neaton - 0000-0001-7585-6135
Harry A. Atwater - 0000-0001-9435-0201

**References:**

(1) Wang, G.; Chernikov, A.; Glazov, M. M.; Heinz, T. F.; Marie, X.; Amand, T.; Urbaszek, B. Colloquium: Excitons in Atomically Thin Transition Metal Dichalcogenides. *Rev Mod Phys* **2018**, *90* (2), 021001.

(2) Xu, Y.; Liu, S.; Rhodes, D. A.; Watanabe, K.; Taniguchi, T.; Hone, J.; Elser, V.; Mak, K. F.; Shan, J. Correlated Insulating States at Fractional Fillings of Moiré Superlattices. *Nature 2020 587:7833* **2020**, *587* (7833), 214–218.

(3) Zhang, Z.; Regan, E. C.; Wang, D.; Zhao, W.; Wang, S.; Sayyad, M.; Yumigeta, K.; Watanabe, K.; Taniguchi, T.; Tongay, S.; Crommie, M.; Zettl, A.; Zaletel, M. P.; Wang, F. Correlated Interlayer Exciton Insulator in Heterostructures of Monolayer WSe2 and Moiré WS2/WSe2. *Nature Physics 2022 18:10* **2022**, *18* (10), 1214–1220.

(4) Smoleński, T.; Dolgirev, P. E.; Kuhlenkamp, C.; Popert, A.; Shimazaki, Y.; Back, P.; Lu, X.; Kroner, M.; Watanabe, K.; Taniguchi, T.; Esterlis, I.; Demler, E.; Imamoğlu, A. Signatures of Wigner Crystal of Electrons in a Monolayer Semiconductor. *Nature 2021 595:7865* **2021**, *595* (7865), 53–57.

(5) Wang, X.; Xiao, C.; Park, H.; Zhu, J.; Wang, C.; Taniguchi, T.; Watanabe, K.; Yan, J.; Xiao, D.; Gamelin, D. R.; Yao, W.; Xu, X. Light-Induced Ferromagnetism in Moiré Superlattices. *Nature 2022 604:7906* **2022**, *604* (7906), 468–473.

(6) Zhu, B.; Chen, X.; Cui, X. Exciton Binding Energy of Monolayer WS2. *Scientific Reports 2015 5:1* **2015**, *5* (1), 1–5.

(7) Qiu, D. Y.; da Jornada, F. H.; Louie, S. G. Optical Spectrum of MoS2: Many-Body Effects and Diversity of Exciton States. *Phys Rev Lett* **2013**, *111* (21), 216805.

(8) Klots, A. R.; Newaz, A. K. M.; Wang, B.; Prasai, D.; Krzyzanowska, H.; Lin, J.; Caudel, D.; Ghimire, N. J.; Yan, J.; Ivanov, B. L.; Velizhanin, K. A.; Burger, A.; Mandrus, D. G.; Tolk, N. H.; Pantelides, S. T.; Bolotin, K. I. Probing Excitonic States in Suspended Two-Dimensional Semiconductors by Photocurrent Spectroscopy. *Scientific Reports 2014 4:1* **2014**, *4* (1), 1–7.

(9) Ugeda, M. M.; Bradley, A. J.; Shi, S. F.; da Jornada, F. H.; Zhang, Y.; Qiu, D. Y.; Ruan, W.; Mo, S. K.; Hussain, Z.; Shen, Z. X.; Wang, F.; Louie, S. G.; Crommie, M. F. Giant Bandgap Renormalization and Excitonic Effects in a Monolayer Transition Metal Dichalcogenide Semiconductor. *Nature Materials 2014 13:12* **2014**, *13* (12), 1091–1095.

(10) Chernikov, A.; Berkelbach, T. C.; Hill, H. M.; Rigosi, A.; Li, Y.; Aslan, O. B.; Reichman, D. R.; Hybertsen, M. S.; Heinz, T. F. Exciton Binding Energy and Nonhydrogenic Rydberg Series in Monolayer WS2. *Phys Rev Lett* **2014**, *113* (7), 076802.

(11) Raja, A.; Chaves, A.; Yu, J.; Arefe, G.; Hill, H. M.; Rigosi, A. F.; Berkelbach, T. C.; Nagler, P.; Schüller, C.; Korn, T.; Nuckolls, C.; Hone, J.; Brus, L. E.; Heinz, T. F.; Reichman, D. R.; Chernikov, A. Coulomb Engineering of the Bandgap and Excitons in Two-Dimensional Materials. *Nature Communications 2017 8:1* **2017**, *8* (1), 1–7.

(12) Xu, X.; Yao, W.; Xiao, D.; Heinz, T. F. Spin and Pseudospins in Layered Transition Metal Dichalcogenides. *Nature Physics 2014 10:5* **2014**, *10* (5), 343–350.

(13) Xiao, D.; Liu, G. bin; Feng, W.; Xu, X.; Yao, W. Coupled Spin and Valley Physics in Monolayers of MoS 2 and Other Group-VI Dichalcogenides. *Phys Rev Lett* **2012**, *108* (19), 196802.

(14) Cao, T.; Wang, G.; Han, W.; Ye, H.; Zhu, C.; Shi, J.; Niu, Q.; Tan, P.; Wang, E.; Liu, B.; Feng, J. Valley-Selective Circular Dichroism of Monolayer Molybdenum Disulphide. *Nature Communications 2012 3:1* **2012**, *3* (1), 1–5.

(15) Diware, M. S.; Ganorkar, S. P.; Park, K.; Chegal, W.; Cho, H. M.; Cho, Y. J.; Kim, Y. D.; Kim, H. Dielectric Function, Critical Points, and Rydberg Exciton Series of WSe2 Monolayer. *Journal of Physics: Condensed Matter* **2018**, *30* (23), 235701.

(16) Chen, S. Y.; Lu, Z.; Goldstein, T.; Tong, J.; Chaves, A.; Kunstmann, J.; Cavalcante, L. S. R.; Woźniak, T.; Seifert, G.; Reichman, D. R.; Taniguchi, T.; Watanabe, K.; Smirnov, D.; Yan, J. Luminescent Emission of Excited Rydberg Excitons from Monolayer WSe 2. *Nano Lett* **2019**, *19* (4), 2464–2471.

(17) Chernikov, A.; Berkelbach, T. C.; Hill, H. M.; Rigosi, A.; Li, Y.; Aslan, O. B.; Reichman, D. R.; Hybertsen, M. S.; Heinz, T. F. Exciton Binding Energy and Nonhydrogenic Rydberg Series in Monolayer WS2. *Phys Rev Lett* **2014**, *113* (7), 076802.

(18) Qiu, D. Y.; da Jornada, F. H.; Louie, S. G. Screening and Many-Body Effects in Two-Dimensional Crystals: Monolayer MoS2. *Phys Rev B* **2016**, *93* (23), 235435.

(19) Popert, A.; Shimazaki, Y.; Kroner, M.; Watanabe, K.; Taniguchi, T.; Imamoğlu, A.; Smoleński, T. Optical Sensing of Fractional Quantum Hall Effect in Graphene. *Nano Lett* **2022**, *22* (18), 7363–7369.

(20) Xu, Y.; Horn, C.; Zhu, J.; Tang, Y.; Ma, L.; Li, L.; Liu, S.; Watanabe, K.; Taniguchi, T.; Hone, J. C.; Shan, J.; Mak, K. F. Creation of Moiré Bands in a Monolayer Semiconductor by Spatially Periodic Dielectric Screening. *Nature Materials 2021 20:5* **2021**, *20* (5), 645–649.

(21) Xu, Y.; Liu, S.; Rhodes, D. A.; Watanabe, K.; Taniguchi, T.; Hone, J.; Elser, V.; Mak, K. F.; Shan, J. Correlated Insulating States at Fractional Fillings of Moiré Superlattices. *Nature 2020 587:7833* **2020**, *587* (7833), 214–218.

(22) Walther, V.; Johne, R.; Pohl, T. Giant Optical Nonlinearities from Rydberg Excitons in Semiconductor Microcavities. *Nature Communications 2018 9:1* **2018**, *9* (1), 1–6.

(23) Gu, J.; Walther, V.; Waldecker, L.; Rhodes, D.; Raja, A.; Hone, J. C.; Heinz, T. F.; Kéna-Cohen, S.; Pohl, T.; Menon, V. M. Enhanced Nonlinear Interaction of Polaritons via Excitonic Rydberg States in Monolayer WSe2. *Nature Communications 2021 12:1* **2021**, *12* (1), 1–7.

(24) Chernikov, A.; Berkelbach, T. C.; Hill, H. M.; Rigosi, A.; Li, Y.; Aslan, O. B.; Reichman, D. R.; Hybertsen, M. S.; Heinz, T. F. Exciton Binding Energy and Nonhydrogenic Rydberg Series in Monolayer WS2. *Phys Rev Lett* **2014**, *113* (7), 076802.

(25) Hill, H. M.; Rigosi, A. F.; Roquelet, C.; Chernikov, A.; Berkelbach, T. C.; Reichman, D. R.; Hybertsen, M. S.; Brus, L. E.; Heinz, T. F. Observation of Excitonic Rydberg States in Monolayer MoS2 and WS2 by Photoluminescence Excitation Spectroscopy. *Nano Lett* **2015**, *15* (5), 2992–2997.

(26) Wagner, K.; Wietek, E.; Ziegler, J. D.; Semina, M. A.; Taniguchi, T.; Watanabe, K.; Zipfel, J.; Glazov, M. M.; Chernikov, A. Autoionization and Dressing of Excited Excitons by Free Carriers in Monolayer WSe2. *Phys. Rev. Lett.* **2020**, *125* (26), 267401.

(27) Goldstein, T.; Wu, Y. C.; Chen, S. Y.; Taniguchi, T.; Watanabe, K.; Varga, K.; Yan, J. Ground and Excited State Exciton Polarons in Monolayer MoSe. *J. Chem. Phys.* **2020**, *153* (7), 071101.

(28) Liu, E.; van Baren, J.; Lu, Z.; Taniguchi, T.; Watanabe, K.; Smirnov, D.; Chang, Y. C.; Lui, C. H. Exciton-Polaron Rydberg States in Monolayer MoSe2 and WSe2. *Nature Communications 2021 12:1* **2021**, *12* (1), 1–8.

(29) Arora, A.; Deilmann, T.; Reichenauer, T.; Kern, J.; Michaelis De Vasconcellos, S.; Rohlfing, M.; Bratschitsch, R. Excited-State Trions in Monolayer WS2. *Phys Rev Lett* **2019**, *123* (16), 167401.

(30) Han, B.; Robert, C.; Courtade, E.; Manca, M.; Shree, S.; Amand, T.; Renucci, P.; Taniguchi, T.; Watanabe, K.; Marie, X.; Golub, L. E.; Glazov, M. M.; Urbaszek, B. Exciton States in Monolayer MoSe2 and MoTe2 Probed by Upconversion Spectroscopy. *Phys Rev X* **2018**, *8* (3), 031073.

(31) Yang, J.; Lü, T.; Myint, Y. W.; Pei, J.; Macdonald, D.; Zheng, J. C.; Lu, Y. Robust Excitons and Trions in Monolayer MoTe2. *ACS Nano* **2015**, *9* (6), 6603–6609.

(32) Chen, B.; Sahin, H.; Suslu, A.; Ding, L.; Bertoni, M. I.; Peeters, F. M.; Tongay, S. Environmental Changes in MoTe2 Excitonic Dynamics by Defects-Activated Molecular Interaction. *ACS Nano* **2015**, *9* (5), 5326–5332.

(33) Froehlicher, G.; Lorchat, E.; Berciaud, S. Direct versus Indirect Band Gap Emission and Exciton-Exciton Annihilation in Atomically Thin Molybdenum Ditelluride (MoTe2). *Phys Rev B* **2016**, *94* (8), 085429.

(34) Helmrich, S.; Schneider, R.; Achtstein, A. W.; Arora, A.; Herzog, B.; de Vasconcellos, S. M.; Kolarczik, M.; Schöps, O.; Bratschitsch, R.; Woggon, U.; Owschimikow, N. Exciton–Phonon Coupling in Mono- and Bilayer MoTe2. *2d Mater* **2018**, *5* (4), 045007.

(35) Ruppert, C.; Aslan, O. B.; Heinz, T. F. Optical Properties and Band Gap of Single- and Few-Layer MoTe2 Crystals. *Nano Lett* **2014**, *14* (11), 6231–6236.

(36) Lezama, I. G.; Arora, A.; Ubaldini, A.; Barreteau, C.; Giannini, E.; Potemski, M.; Morpurgo, A. F. Indirect-to-Direct Band Gap Crossover in Few-Layer MoTe2. *Nano Lett* **2015**, *15* (4), 2336–2342.

(37) Arora, A.; Schmidt, R.; Schneider, R.; Molas, M. R.; Breslavetz, I.; Potemski, M.; Bratschitsch, R. Valley Zeeman Splitting and Valley Polarization of Neutral and Charged Excitons in Monolayer MoTe2 at High Magnetic Fields. *Nano Lett* **2016**, *16* (6), 3624–3629.

(38) Koirala, S.; Mouri, S.; Miyauchi, Y.; Matsuda, K. Homogeneous Linewidth Broadening and Exciton Dephasing Mechanism in MoT E2. *Phys Rev B* **2016**, *93* (7), 075411.

(39) Wang, Y.; Xiao, J.; Zhu, H.; Li, Y.; Alsaid, Y.; Fong, K. Y.; Zhou, Y.; Wang, S.; Shi, W.; Wang, Y.; Zettl, A.; Reed, E. J.; Zhang, X. Structural Phase Transition in Monolayer MoTe2 Driven by Electrostatic Doping. *Nature 2017 550:7677* **2017**, *550* (7677), 487–491.

(40) Duerloo, K. A. N.; Li, Y.; Reed, E. J. Structural Phase Transitions in Two-Dimensional Mo- and W-Dichalcogenide Monolayers. *Nature Communications 2014 5:1* **2014**, *5* (1), 1–9.

(41) Sidler, M.; Back, P.; Cotlet, O.; Srivastava, A.; Fink, T.; Kroner, M.; Demler, E.; Imamoglu, A. Fermi Polaron-Polaritons in Charge-Tunable Atomically Thin Semiconductors. *Nat. Phys.* **2017**, *13* (3), 255.

(42) Fante, R. L.; Mccormack, M. T. Reflection Properties of the Salisbury Screen. *IEEE Trans Antennas Propag* **1988**, *36* (10), 1443–1454.

(43) Lu, Y. J.; Sokhoyan, R.; Cheng, W. H.; Kafaie Shirmanesh, G.; Davoyan, A. R.; Pala, R. A.; Thyagarajan, K.; Atwater, H. A. Dynamically Controlled Purcell Enhancement of Visible Spontaneous Emission in a Gated Plasmonic Heterostructure. *Nature Communications 2017 8:1* **2017**, *8* (1), 1–8.

(44) Berghäuser, G.; Malic, E. Analytical Approach to Excitonic Properties of MoS 2. *Phys Rev B Condens Matter Mater Phys* **2014**, *89* (12), 125309.

(45) Qiu, D. Y.; da Jornada, F. H.; Louie, S. G. Environmental Screening Effects in 2D Materials: Renormalization of the Bandgap, Electronic Structure, and Optical Spectra of Few-Layer Black Phosphorus. *Nano Lett* **2017**, *17* (8), 4706–4712.

(46) Li, L.; Kim, J.; Jin, C.; Ye, G. J.; Qiu, D. Y.; da Jornada, F. H.; Shi, Z.; Chen, L.; Zhang, Z.; Yang, F.; Watanabe, K.; Taniguchi, T.; Ren, W.; Louie, S. G.; Chen, X. H.; Zhang, Y.; Wang, F. Direct Observation of the Layer-Dependent Electronic Structure in Phosphorene. *Nature Nanotechnology 2016 12:1* **2016**, *12* (1), 21–25.

(47) Bradley, A. J.; Ugeda, M. M.; da Jornada, F. H.; Qiu, D. Y.; Ruan, W.; Zhang, Y.; Wickenburg, S.; Riss, A.; Lu, J.; Mo, S. K.; Hussain, Z.; Shen, Z. X.; Louie, S. G.; Crommie, M. F. Probing the Role of Interlayer Coupling and Coulomb Interactions on Electronic Structure in Few-Layer MoSe2 Nanostructures. *Nano Lett* **2015**, *15* (4), 2594–2599.

(48) Liang, Y.; Yang, L. Carrier Plasmon Induced Nonlinear Band Gap Renormalization in Two-Dimensional Semiconductors. *Phys Rev Lett* **2015**, *114* (6), 063001.

(49) Gao, S.; Liang, Y.; Spataru, C. D.; Yang, L. Dynamical Excitonic Effects in Doped Two-Dimensional Semiconductors. *Nano Lett* **2016**, *16* (9), 5568–5573.

(50) Katsch, F.; Knorr, A. Excitonic Theory of Doping-Dependent Optical Response in Atomically Thin Semiconductors. *Phys Rev B* **2022**, *105* (4), 045301.

(51) Brem, S.; Zipfel, J.; Selig, M.; Raja, A.; Waldecker, L.; Ziegler, J. D.; Taniguchi, T.; Watanabe, K.; Chernikov, A.; Malic, E. Intrinsic Lifetime of Higher Excitonic States in Tungsten Diselenide Monolayers. *Nanoscale* **2019**, *11* (25), 12381–12387.

(52) Palummo, M.; Bernardi, M.; Grossman, J. C. Exciton Radiative Lifetimes in Two-Dimensional Transition Metal Dichalcogenides. *Nano Lett* **2015**, *15* (5), 2794–2800.

**Figures:**

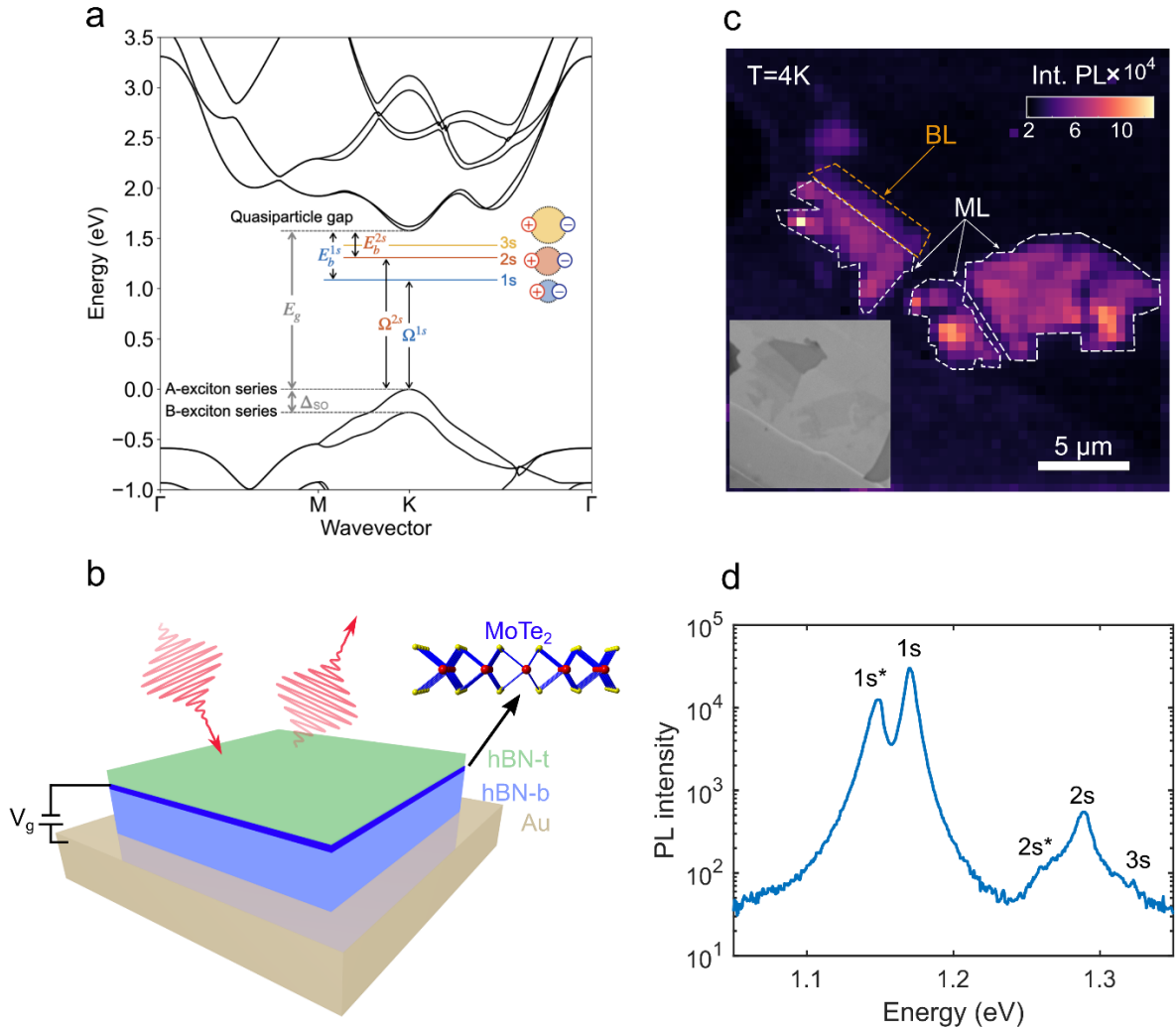

Figure 1. Electro-optic investigation of Rydberg excitons in monolayer MoTe$_2$. (a) Excitonic energy landscape of Rydberg series in monolayer MoTe$_2$ with the quasiparticle band structure, exciton state energies $\Omega^S$, and exciton binding energies $E_b^S$ obtained using GW-BSE calculations. (b) Investigated device geometry consisting of hBN encapsulated monolayer MoTe$_2$ on Au substrate with applied gate voltage. (c) Integrated PL intensity map of investigated sample at 4K. Bright spots indicate monolayer. Inset – optical micrograph of sample. (d) Example PL spectra with assigned Rydberg states.

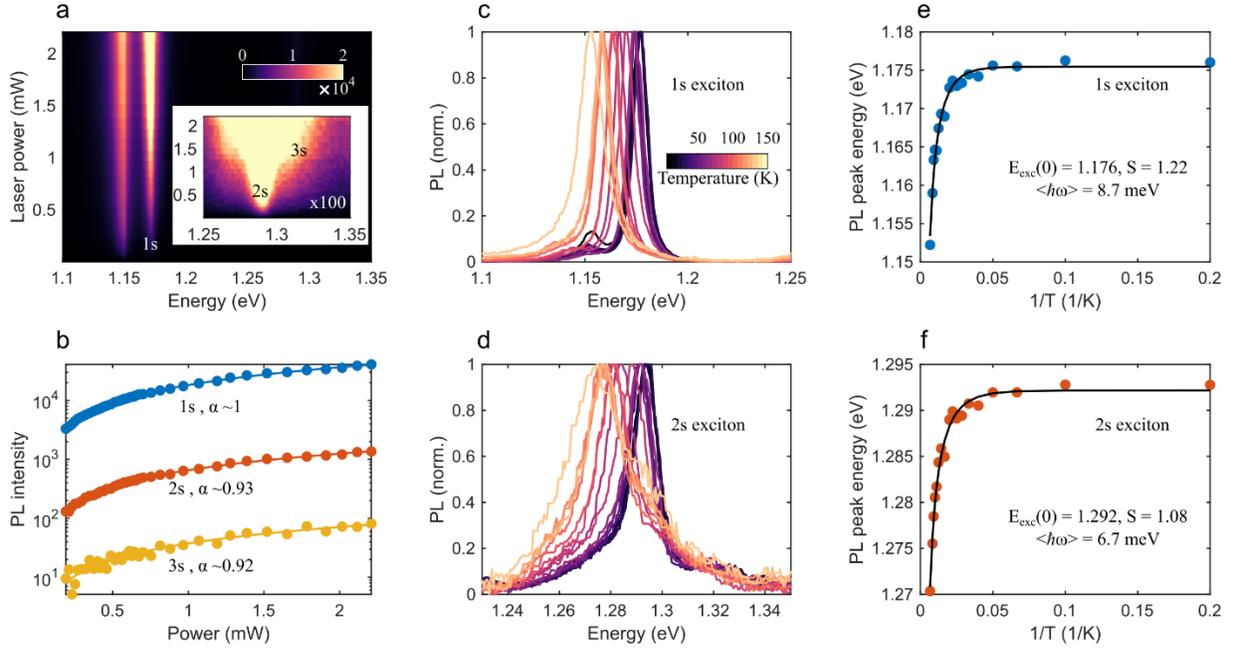

Figure 2. Pump power and temperature dependence of Rydberg excitons. (a) PL intensity variation with increasing pump power for the 1s exciton. Inset – PL(x100) for the 2s and 3s exciton. (b) Semi-log scale plot of intensity dependence of PL with pump power and corresponding fits to a power-law showing excitonic emission ($I_{PL} = I_0\, P^{\alpha}$). (c), (d) Temperature variation of normalized PL spectrum for the 1s and 2s exciton regions, respectively. (e), (f) Fits to a temperature model estimating different parameters for the 1s exciton and 2s exciton.

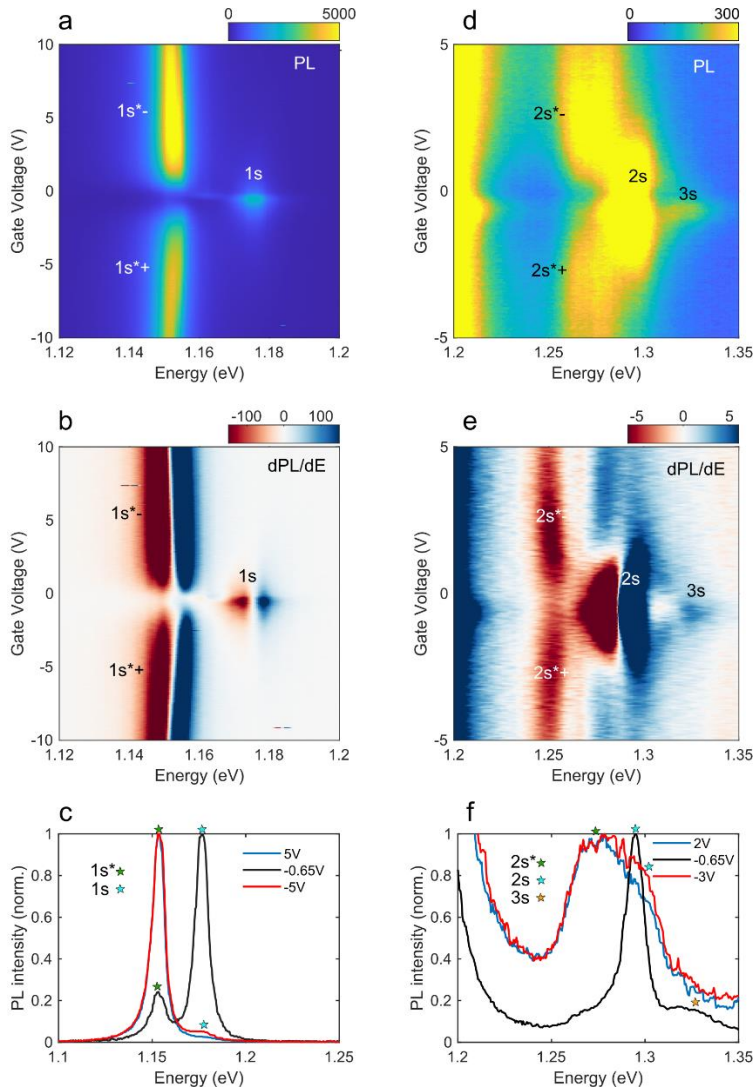

Figure 3. Gate dependent PL spectrum of Rydberg excitons. (a), (d) PL intensity of different neutral exciton species and their corresponding trion features as a function of gate voltage near the 1s and 2s/3s resonance, respectively. (b), (e) Derivative of the PL spectra, $\frac{dPL}{dE}$, shown in (a), (d). (c), (f) Line cuts of the PL spectrum at different voltages showing the different exciton and trion resonances.

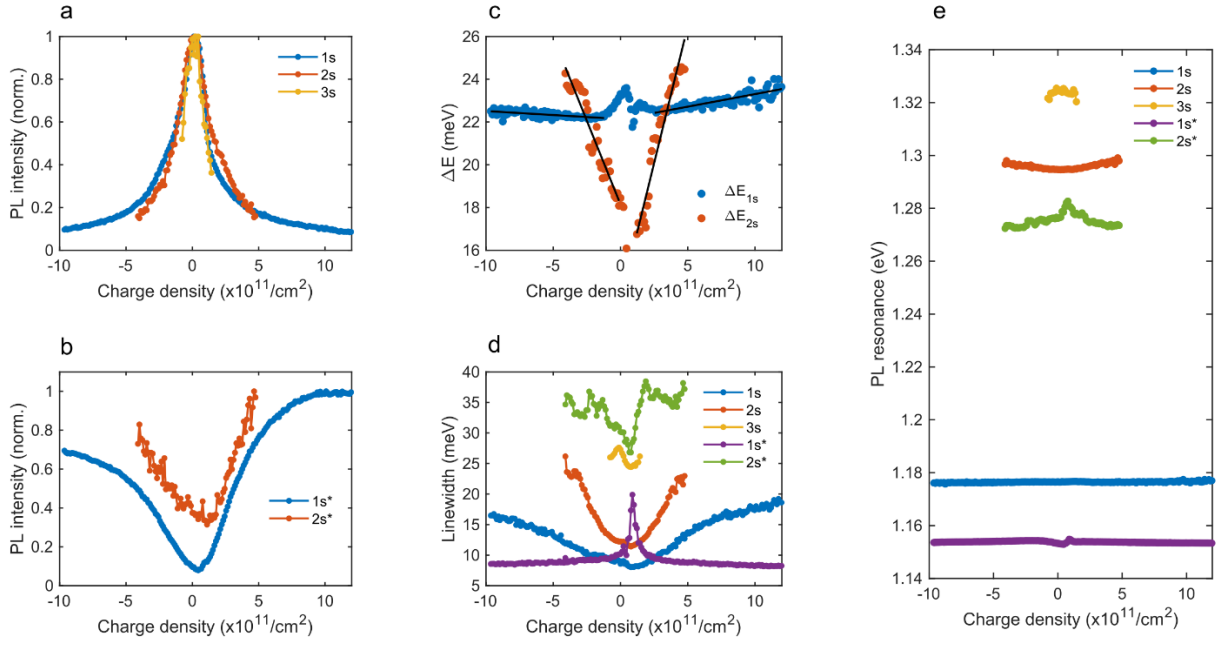

Figure 4. Gate tunable PL properties of Rydberg excitons. (a), (b) PL intensity (normalized) of different neutral exciton, trion species, respectively, as a function of charge density. (c) Energy shifts between the neutral exciton and the trion for 1s and 2s states as a function of charge density. (d) Evolution of the linewidth and (e) resonance energy of various exciton and trion states as a function of charge density.

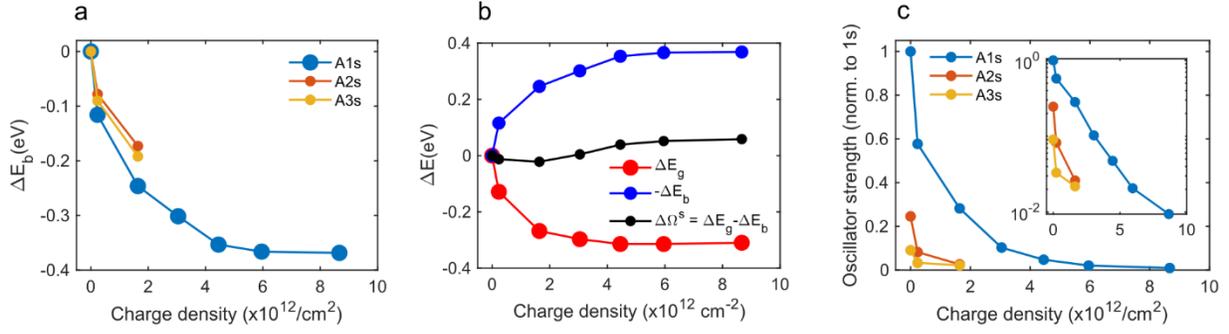

Figure 5. Doping dependence of (a) the variation in exciton binding energy, $\Delta E_b$, for the ground A1s exciton, and the excited A2s and A3s states, (b) the variation in exciton energy $\Delta \Omega^s$ (black curve), the exciton binding energy $\Delta E_b$ (blue curve), and the renormalization of the QP band gap $\Delta E_g$ (red curve) for the ground A1s exciton, (c) the oscillator strength for the A1s, A2s and A3s states. (inset) Same as (c) plotted in semi-log scale on the y-axis.

# Rydberg Excitons and Trions in Monolayer MoTe$_2$

Souvik Biswas[1,11], Aurélie Champagne[2,3], Jonah B. Haber[3], Supavit Pokawanvit[4,5], Joeson Wong[1,11], Hamidreza Akbari[1], Sergiy Krylyuk[6], Kenji Watanabe[7], Takashi Taniguchi[8], Albert V. Davydov[6], Zakaria Y. Al Balushi[1], Diana Y. Qiu[9], Felipe H. da Jornada[5], Jeffrey B. Neaton[2,3,10], Harry A. Atwater[1,11]

Affiliations
1. Thomas J. Watson Laboratory of Applied Physics, California Institute of Technology, Pasadena, CA 91125, USA
2. Materials and Chemical Science Division, Lawrence Berkeley National Laboratory, Berkeley, California 94720, USA
3. Department of Physics, University of California Berkeley, Berkeley, California 94720, USA
4. Department of Applied Physics, Stanford University, Stanford, CA 94305, USA
5. Department of Materials Science and Engineering, Stanford University, Stanford, California 94305, USA.
6. Materials Science and Engineering Division, National Institute of Standards and Technology, Gaithersburg, Maryland 20899, USA
7. Research Center for Functional Materials, National Institute for Materials Science, 1-1Namiki, Tsukuba 305-0044, Japan
8. International Center for Materials, Nanoarchitectonics, National Institute for Materials Science, 1-1 Namiki, Tsukuba 305-0044, Japan
9. Department of Mechanical Engineering and Materials Science, Yale University, New Haven, Connecticut 06520, United States
10. Kavli Energy Nanosciences Institute at Berkeley, Berkeley, CA 94720, USA
11. Kavli Nanoscience Institute Pasadena, CA 91125, USA

## Supplementary Information

**Table of contents**

U. Atomic force microscope image of MoTe$_2$ device

## A. Crystal growth, device fabrication, and experimental methods.

SiO$_2$ (285 nm)/Si chips were cleaned with ultrasonication in acetone and isopropanol for 30 min. each, followed by oxygen plasma treatment at 70 W, 300 mTorr for 5 min. Monolayer MoTe$_2$, few layer hBN and graphene flakes were directly exfoliated using Scotch-tape at 100 °C to increase the yield of monolayer flakes. Monolayer thickness was initially identified using optical contrast (~7 % contrast per layer) and later verified with atomic force microscopy. hBN thickness was confirmed with atomic force microscopy. Flakes were assembled with a polycarbonate/polydimethylsiloxane (PC/PDMS) stamp with pick-up at temperatures between 80 °C -110 °C. The entire heterostructure stack was dropped on prefabricated gates at 180 °C. The polymer was washed off in chloroform overnight, followed by isopropanol for 10 min. Given the air-sensitive nature of MoTe$_2$ monolayers, exfoliation, identification of MoTe$_2$ flakes and stacking of the entire heterostructures were done in a nitrogen purged glovebox with oxygen and moisture levels below 0.5 ppm. Gates were fabricated on SiO$_2$ (285 nm)/Si chips with electron beam lithography (PMMA 950 A4 spun at 3500 rpm for 1 min. and baked at 180 °C, 10 nA beam current and dosage of 1350 $\mu$C/cm$^2$ at 100 kV), developing in methyl isobutyl ketone: isopropanol (1:3) for 1 min., followed by isopropanol for 30 s and electron beam evaporating 5 nm Ti/95 nm Au at 0.5 Å/$s$ deposition rate. Liftoff was done in warm acetone (60°C) for 10 min., followed by rinse in isopropanol for 5 min. The gates were precleaned before drop-down of heterostructure by annealing in high vacuum (2x10$^{-7}$ Torr) at 300 °C for 6 h. The chip was then wire-bonded with Aluminum wires on to a custom home-made printed circuit board.

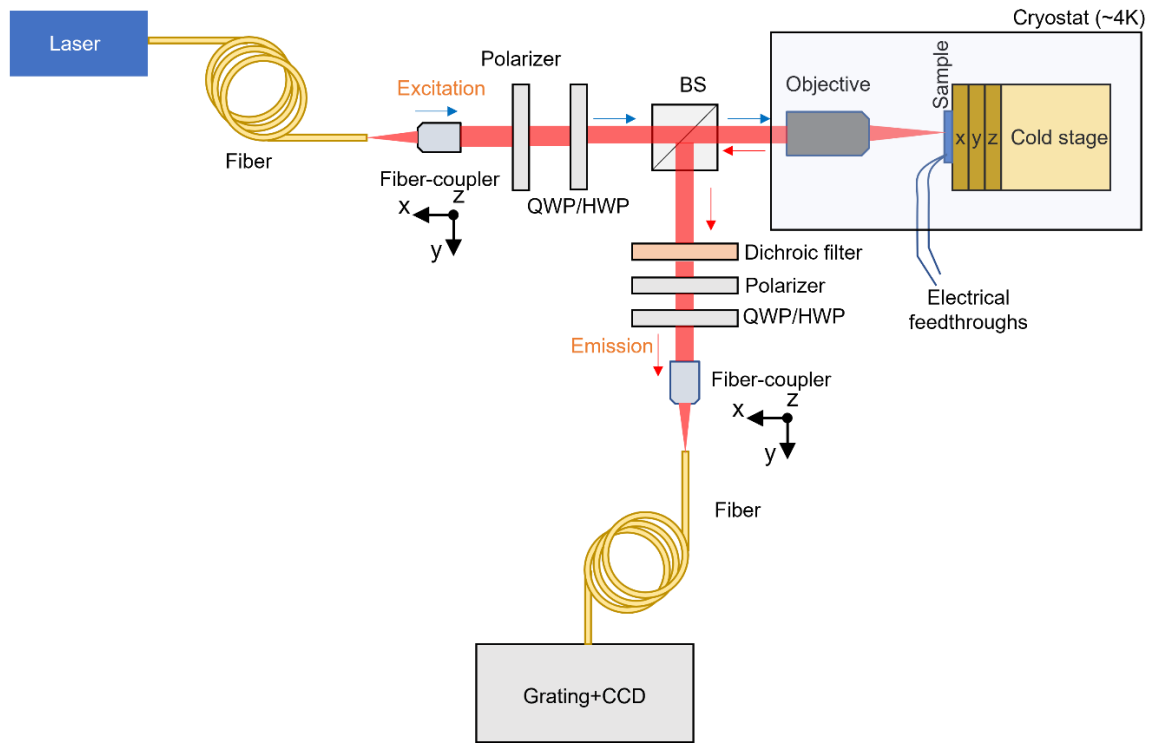

Figure A1. Schematic of the low-temperature confocal optical PL setup used. BS – Beam Splitter, QWP/HWP – Quarter/Half Wave Plate.

Low temperature confocal photoluminescence measurements were performed in an attoDRY800 closed-cycle cryostat at base pressures of <2x10$^{-5}$ mbar. Sample was mounted on a thermally conducting stage with Apiezon glue and stage was cooled using a closed-cycle circulating liquid helium loop. Temperature was varied between ~4K and 300K. A 532 nm (Cobolt) continuous wave laser was used as the excitation source with power ranging between 10 nW and 1 mW, focused to a diffraction limited spot. Emission was collected in a confocal fashion with a cryogenic compatible apochromatic objective with an NA of 0.82 (for the visible and NIR range, LT APO VISIR) and dispersed onto a grating-based spectrometer (with 150 grooves per mm with a Silicon CCD) – Princeton Instruments HRS 300. Voltage was applied using a Keithley 2400. Data was acquired with home-written MATLAB codes.

MoTe$_2$ single-crystals were grown by the chlorine-assisted chemical vapor transport (CVT) method. A vacuum-sealed quartz ampoule with polycrystalline MoTe$_2$ powder and a small amount of TeCl$_4$ transport agent (4 mg per cm$^3$ of ampoules' volume) was placed in a furnace containing a temperature gradient so that the MoTe$_2$ charge was kept at 825 °C, and the temperature at the opposite end of the ampoule was about 710 °C. The ampoule was slowly cooled after 6 days of growth. The 2H phase of the obtained MoTe$_2$ flakes was confirmed by powder X-ray diffraction and transmission electron microscopy studies. Flakes obtained from the aforementioned source as well as commercially available MoTe$_2$ (2D Semiconductors) were investigated with similar results.

Atomic Force Microscopy was performed using Bruker Dimension Icon in tapping mode. Data analysis was performed in MATLAB.

## B. Choice of optical geometry.

The optical geometry adopted here is a Salisbury screen resonant near the A1s exciton transition. Such an optical cavity is illustrated in Figure A2. An optical spacer (dielectric) is placed atop a back reflector (metal or dielectric mirror) and an absorbing layer is placed on the spacer. The thickness of the spacer is chosen to be quarter wavelength ($\lambda/4n$) such that after one round trip in the cavity the light has travelled a total distance of $\lambda/2n$ which introduces a $\pi$ phase shift between the incoming and outgoing electromagnetic wave. This enables near perfect light absorption at the resonance wavelength enabling strong light-matter interaction. In our structures, since we have a top hBN the Salisbury screen condition was estimated from a transfer matrix approach to find the right hBN thicknesses which was used as a guide to select suitable hBN top and bottom flakes. The modelling assumed the following parameters –

1. hBN dielectric/optical spacer ($n = 2.2, k = 0$)
2. MoTe$_2$ monolayer absorbing sheet (optical conductivity model)
   $\sigma(\omega) = 4i\sigma_0 p_0 \omega/((\omega - \omega_0) + i\Gamma/2)$, where $\sigma_0 = e^2/4\hbar$, $p_0 = \gamma_r/(4\pi\omega_0\alpha)$
   $\gamma_r$ (2 $meV$) is the oscillator strength, $\omega_0$ (1.17 $eV$) is the resonance frequency and $\Gamma$ (7 $meV$) is the broadening associated with the optical transition.
3. Refractive index of Au (back reflector) was adopted from Johnson and Christy.
4. A top layer of hBN (5 nm) was added.

Calculations were done using a standard 1D transfer matrix model (with stackrt function in Lumerical as well as home-written MATLAB code). Furthermore, our choice of a back-reflector of gold suppresses any photoluminescence coming from the silicon substrate, which would be emitted in a similar wavelength/energy range $\lambda_{Si} \sim 1150 nm, E_{Si} \sim 1.07\ eV$ at low temperatures.

In a false color map of absorption spectrum variation with bottom hBN thickness shows high absorption at the excitonic resonance for thicknesses around 100 nm. Motivated by these calculations, the bottom hBN for the device was chosen to be 97.3 nm (confirmed by AFM measurements) – limited by the occurrence of naturally exfoliated thicknesses.

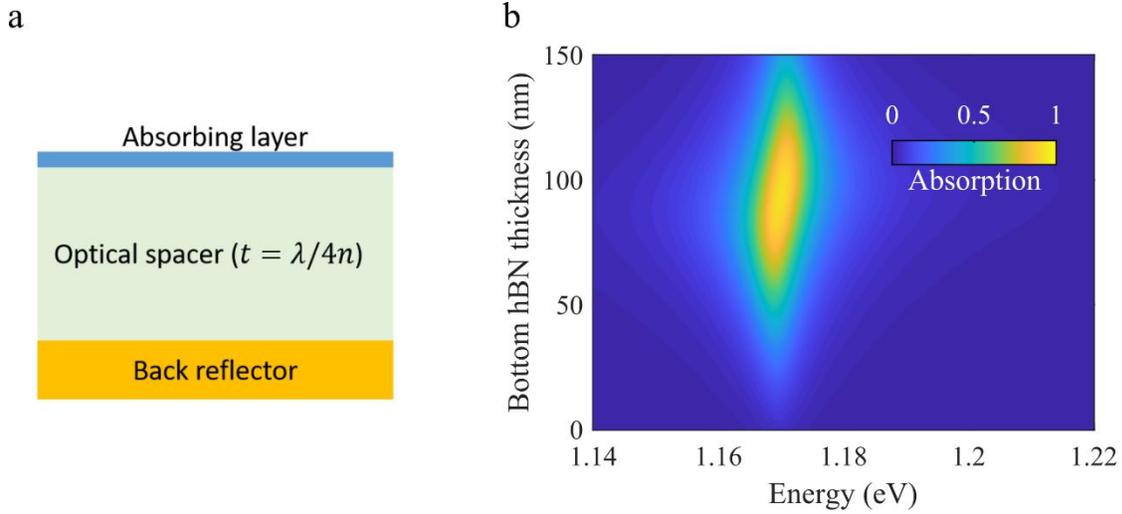

Figure A2. (a) Schematic of a quarter wavelength Salisbury screen geometry. (b) Absorption spectrum as a function of bottom hBN thickness showing the cavity-enhancement.

## C. Estimation of Purcell enhancement

Simulation set-up details – Initially, the transfer matrix (1D) model was used to optimize the structure for a resonant Salisbury screen design. The obtained parameters were then fed into Lumerical FDTD (schematic in Figure A3a) as follows –

- hBN dielectric/optical spacer ($n = 2.2, k = 0$), total thickness of 102.3 nm.
- Refractive index of Au (back reflector) was adopted from Johnson and Christy, assumed to be optically thick (~150 nm).
- A uniform mesh of 2nm was used in all directions around the dipole and the total simulation span was taken to be $5\mu m$, to eliminate any artifacts.

An x-polarized dipole (in-plane of the simulation) was used to mimic the emission from MoTe$_2$. Since the physical thickness of a monolayer ($t \sim 0.7\ nm$) is much smaller than the wavelengths of interest ($\lambda \sim 1\mu m$) it is justified to assume the dipole is completely in plane with no out-of-plane component. Perfectly matched layers (PML) boundary conditions were used on all sides and the dipole wavelength was chosen to match the A1s exciton emission ($\lambda = 1.172\ eV$ or $1058\ nm$). Field profile (power) monitors were placed around the dipole in XZ and YZ configuration and the total field (real) intensity recorded as shown in Figure A3b and c respectively. From the dipole analysis (in built) in Lumerical FDTD, we estimated a Purcell factor - $F_p = 2.012$.

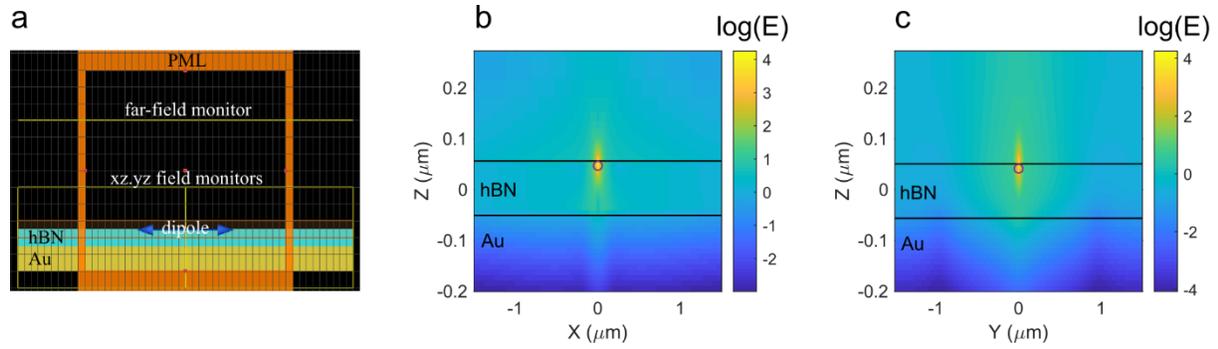

Figure A3. (a) Simulation setup XZ view in Lumerical FDTD. (b), (c) XZ, YZ monitor for log(Re(E)) profile, respectively.

### D. Comparison of monolayer and bilayer PL spectra.

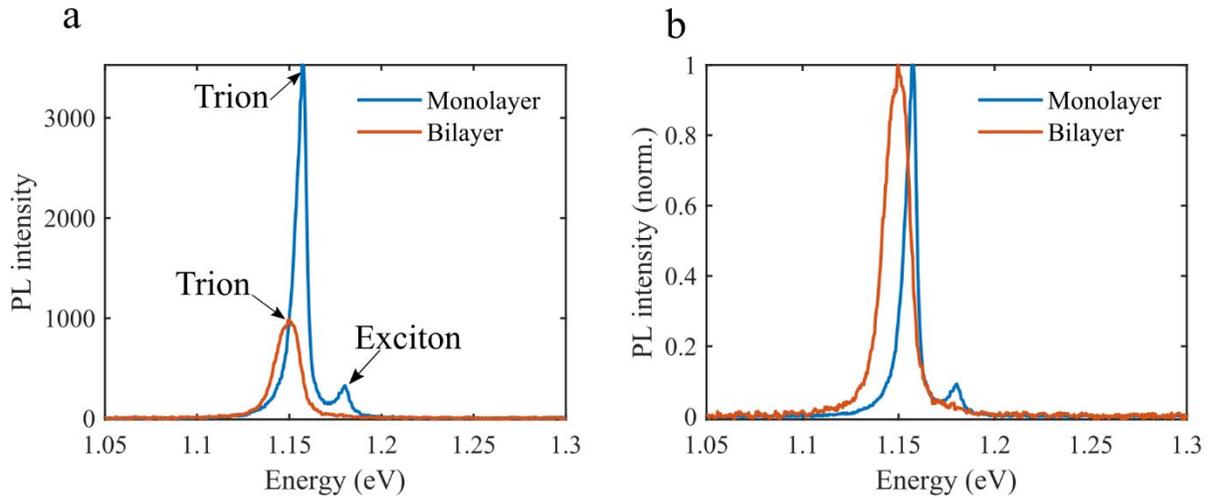

Figure A4. (a) PL spectrum from monolayer and bilayer regions of the same device. (b) Same as (a) but normalized.

The trion peak dominates for both the monolayer (1.157 eV) and bilayer (1.149 eV) spectrum under no applied bias. The exciton peak for the monolayer occurs at ~1.181 eV. The sample is lightly n-doped as confirmed by gate dependent measurements. The trion peak for the bilayer is 8 meV lower in energy than the monolayer and about 3.5 times less bright. Since the doping density is the same for the monolayer and the bilayer as they are part of the same flake, the intensities are linked to the quantum yield differences.

### E. Narrowest linewidth observed for A1s exciton.

The narrowest linewidth obtained from a monolayer region of the device was 4.48 meV for the neutral A1s exciton – as fit to a Lorentzian spectrum.

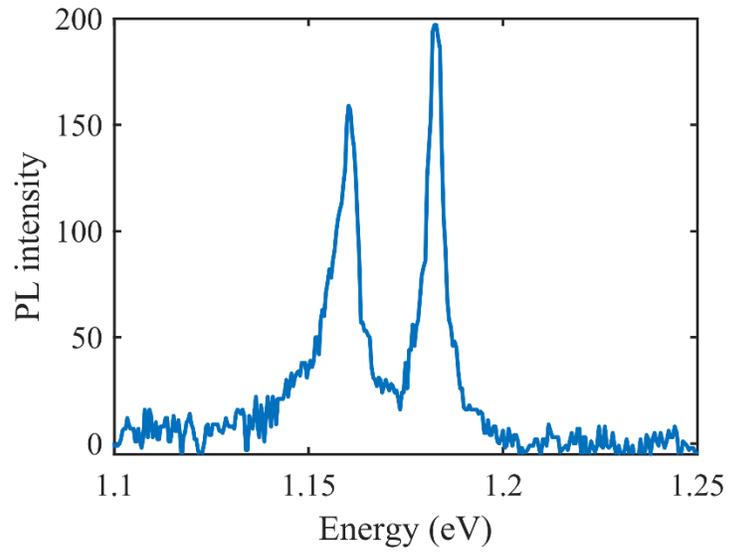

Figure A5. Narrowest emission line obtained for the A1s neutral exciton transition.

### F. Optical image of MoTe$_2$ monolayers.

The following two monolayer (ML) region containing flakes were used in the device fabrication presented in the main text (both show ~7% contrast). A bilayer (BL) region is also seen in the first flake. Scale bar - $2.5\mu m$.

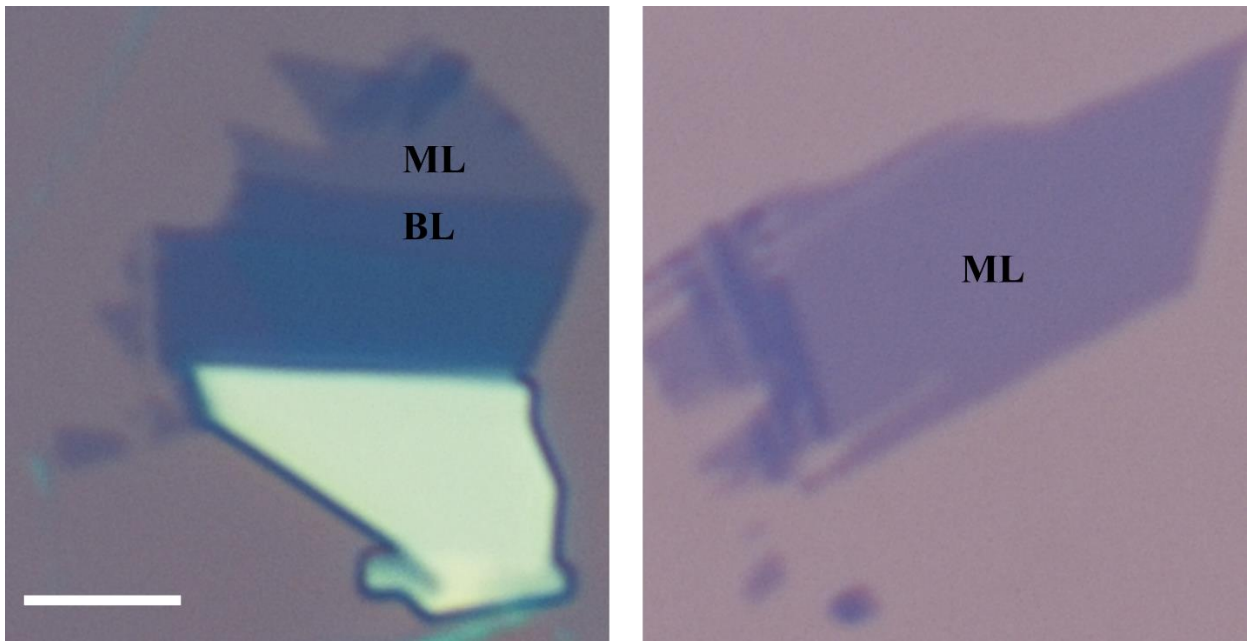

Figure A6. Optical microscope image of flakes used for the device fabrication.

## G. Charge density calculations.

We assume a parallel-plate capacitor model where the two plates of the capacitor are the back reflector/electrode (gold) and the monolayer MoTe$_2$. This assumption is justified in the dc limit since the MoTe2 is semiconducting. This yields the following relation –

$C/A = \epsilon_0 \epsilon_{r,hBN}/d_{hBN}$, where $\epsilon_{r,hBN} = 3.9$ and $d_{hBN} = 97.3 \ nm$.

$n = C(V - V_{CNP}) = \epsilon_0 \epsilon_{r,hBN}(V - V_{CNP})/d_{hBN}$, where $V_{CNP} = -0.65V$ (as estimated from PL measurements).

The relation between applied gate voltage and sheet charge density in MoTe$_2$ is depicted in Figure A7.

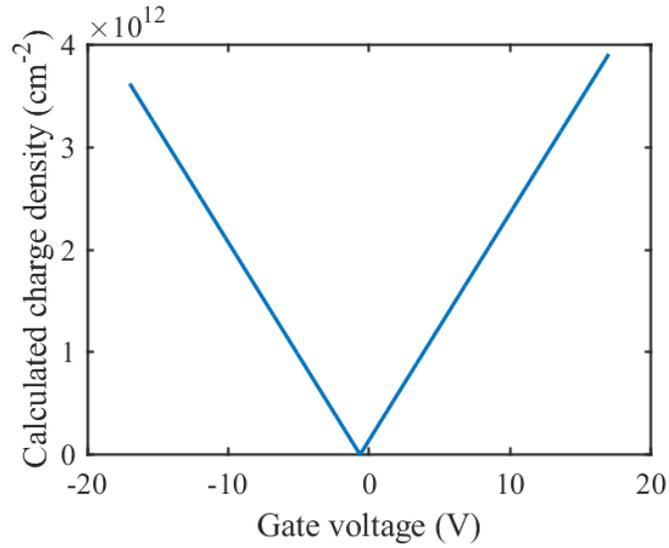

Figure A7. Applied gate voltage to sheet charge density conversion assuming parallel plate capacitor model in the dc limit.

## H. Power dependent emission spectrum.

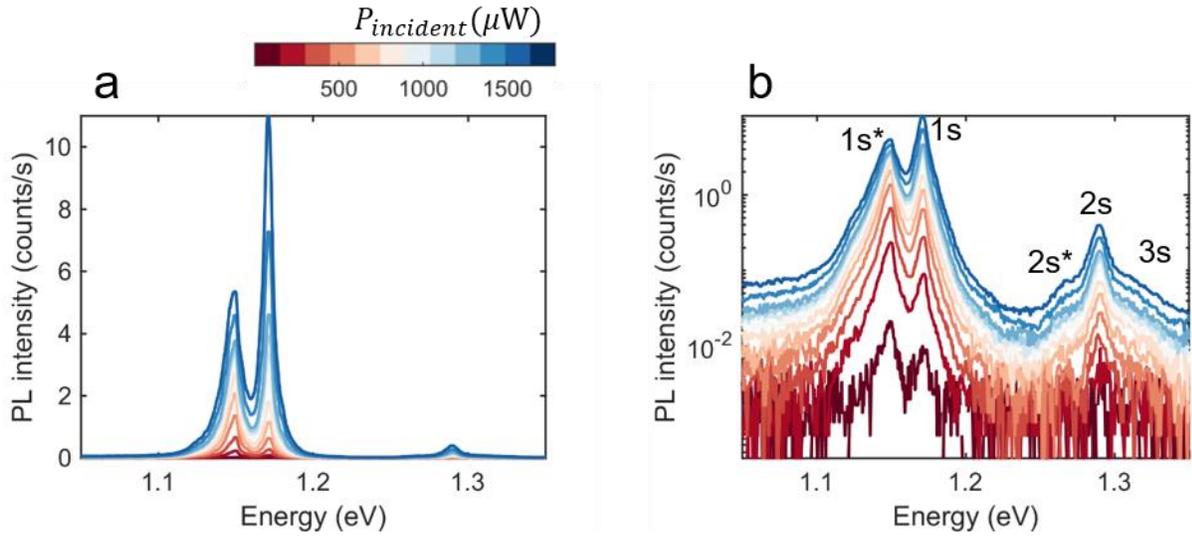

Figure A8. Power dependent photoluminescence spectrum over 3 decades of pump intensity in (a) linear and (b) log scale. The resonances are labelled in (b).

## I. Temperature dependent PL broadening and intensity.

Temperature variation of the PL spectrum shows an increase in the broadening of the linewidth which stems from the phonon contribution. Additionally, the intensity for the 1s drops faster compared to the 2s state.

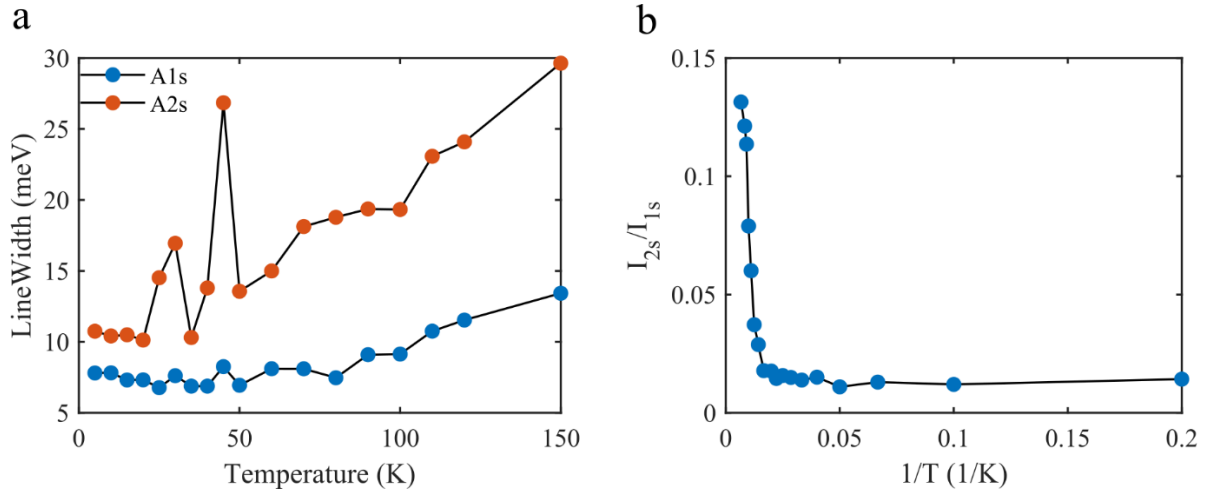

Figure A9. (a) Temperature dependent linewidth broadening of the A1s and A2s excitonic state. (b) Intensity ratio of the 2s/1s excitonic state with temperature (1/T).

## J. Gate dependent PL fits

Fits to the PL spectrum data presented in the main manuscript are shown in Figure A10,11 showing good match between the two.

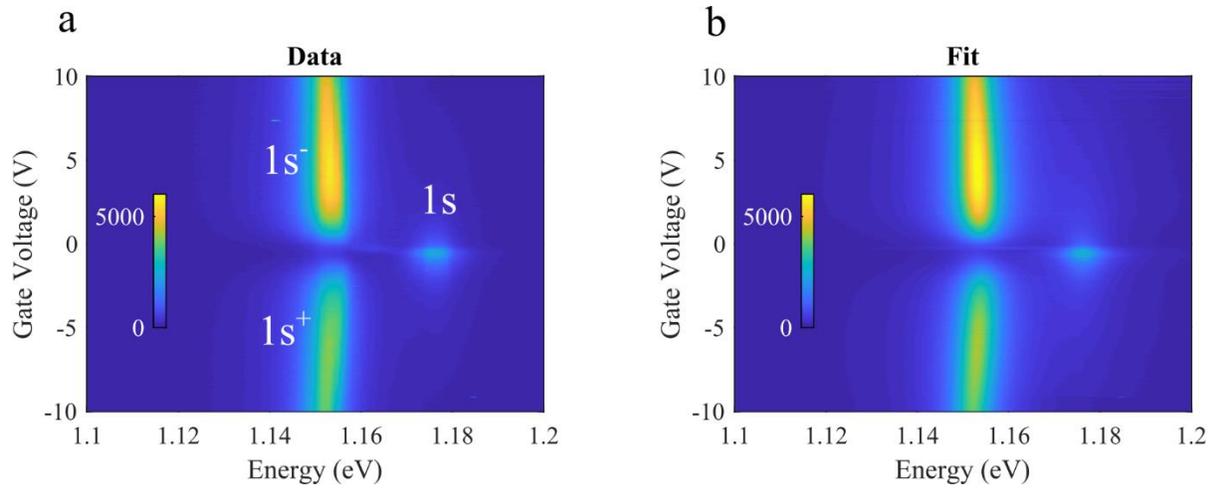

Figure A10. (a) Experimental PL data around the A1s resonance region showing the neutral exciton and the charged trion resonances. (b) Multi-Lorentzian fit to the PL data shown in (a).

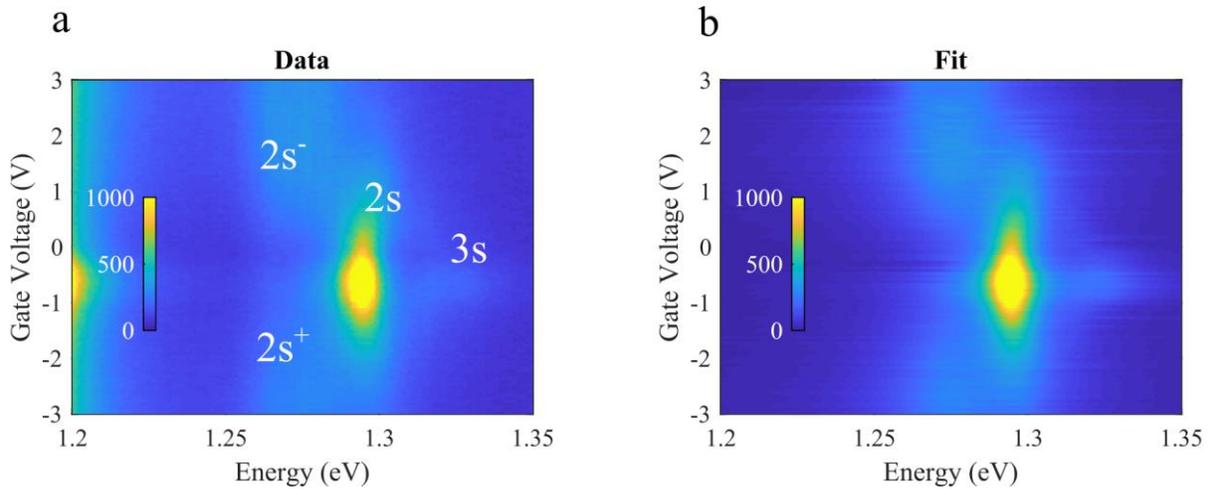

Figure A11. (a) Experimental PL data around the A2s/3s resonance region showing the neutral exciton and the charged trion resonances. (b) Multi-Lorentzian fit to the PL data shown in (a).

## K. Gate dependence of additional spot.

We have measured across 30 spots in the two devices presented in the paper and cycled the same spot through more than 10 sequences of gating between -17V and 17V and also repeated the measurement on a third device and all of them have yielded similar results. The only observable that varied across the spots and samples is the photoluminescence quantum yield which is well-known in the van der Waals community to be inhomogeneous across samples due to imperfections in fabrication techniques resulting in strain and bubble formation in-between layers in heterostructures. Results from another spot are summarized in Figure A12. The resonances have been labelled in the previous sections of the supplementary and the main text.

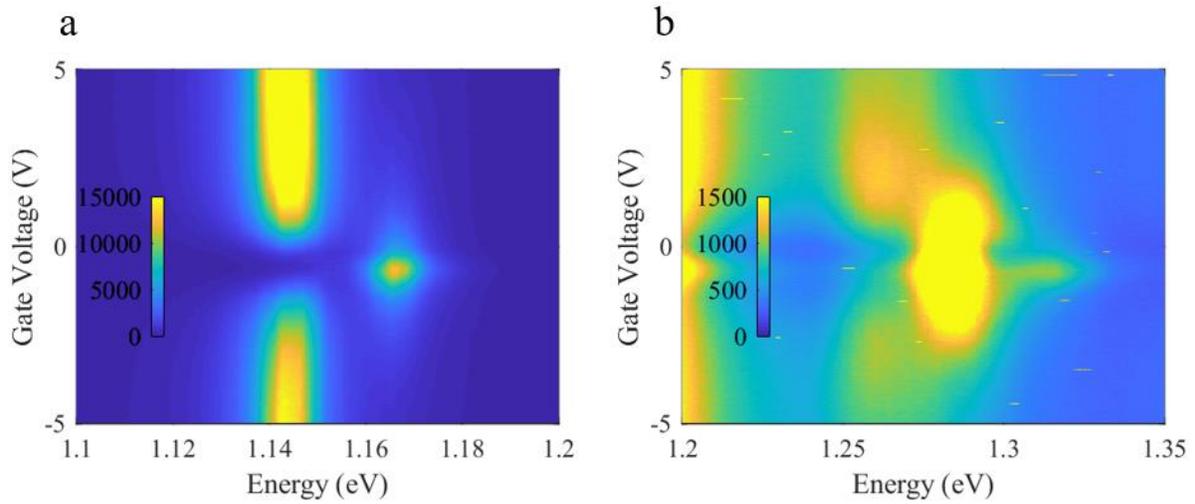

Figure A12. (a) Experimental PL data around the A1s resonance region showing the neutral exciton and the charged trion resonances. (b) Same as (a) but for 2s/3s resonances.

## L. Absolute intensity of exciton and trion emission modulation.

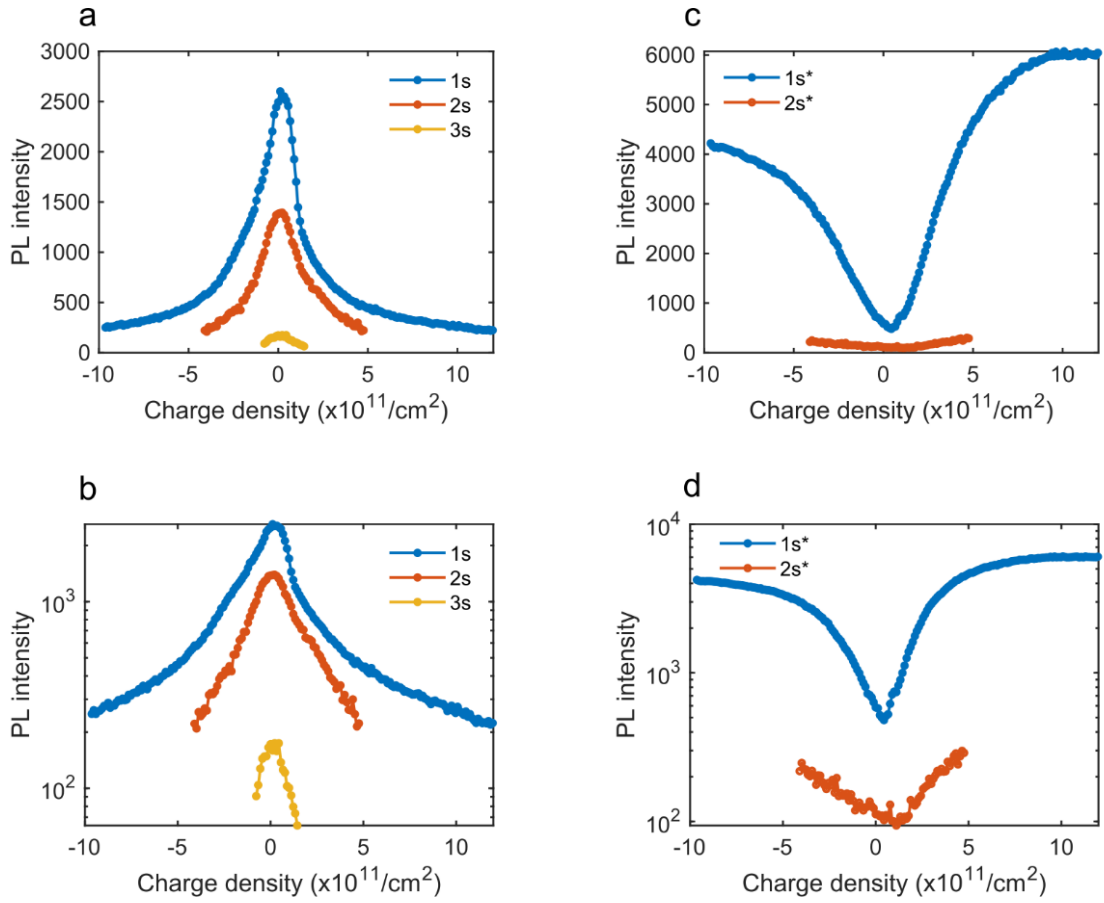

Figure A13. (a) Absolute photoluminescence intensity of Rydberg excitons as a function of charge density in linear scale. (b) Same as (a) in semi-log scale. (c) Absolute photoluminescence intensity of trions associated with Rydberg excitons as a function of charge density in linear scale. (d) Same as (c) in semi-log scale.

## M. Energy shift of exciton and trion emission.

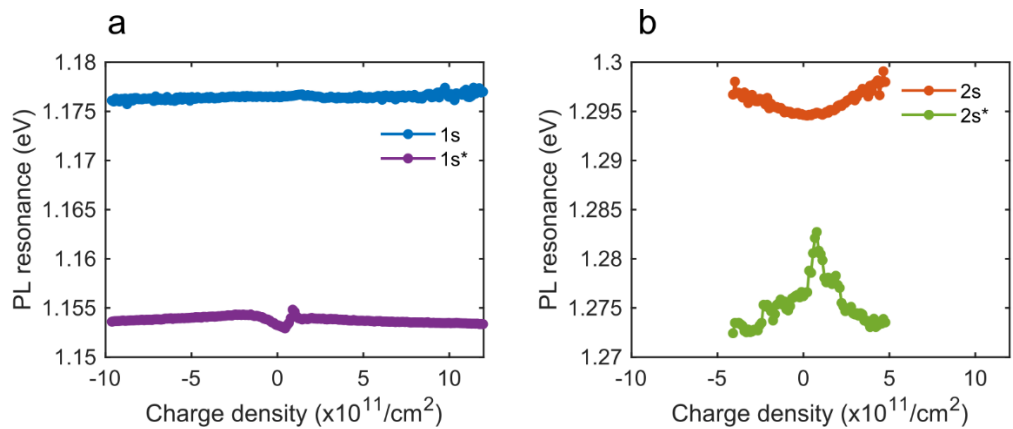

Figure A14. (a) Energy shifts of the A1s exciton and trion as a function of doping density. (b) Energy shifts of the A2s exciton and trion as a function of doping density.

### N. Comparison of MoTe₂ with MoS₂, MoSe₂, WS₂ and WSe₂.

Table TS1 – Binding energy of Rydberg excitons of all TMDCs

| Material | A1s | A1s+ | A1s- | A2s | A2s+ | A2s- | A3s | A3s+ | A3s- |
|---|---|---|---|---|---|---|---|---|---|
| MoS$_2$ | 440(±80)[1], 261[2], 221[3], 222[4] | - | 27[2], 18(±1.5)[5] | 91[2], 48[4] | - | - | 53[2], 20[4] | - | - |
| MoSe$_2$ | 231[3], 208[6] | 24.3(±0.1)[7], 23.7(±0.1)[7], 27[6] | 26.1(±0.1)[7], 23.1(±0.2)[7] | 56[6] | 22.7(±0.4)[7], 22.6(±1.0)[7], 27[6] | 24.6(±0.2)[7], 16.4(±0.7)[7] | | 13.1(±0.3)[7] | 13(±0.5)[7] |
| WS$_2$ | 320(±50)[1], 180[3] | - | - | - | - | - | - | - | - |
| WSe$_2$ | 172[8], 167[3], 170[9], 169[10], 370[11] | 20.5(±0.1)[7], 17.5(0.1)[7] | 29(±0.1)[7], 27.1(±0.1)[7] | 41[8], 39[9], 40[10] | 18.4(±0.3)[7], 9.2(±0.7)[7], 14.1[12] | 17.8(±0.4)[7], 18.6[12] | 20[8], 13[9], 17[10] | - | - |
| MoTe$_2$ | 177[3], 156[13], 580(±80)[14], **404(±19)**, 490 | 24[15], 24[14], **22.14(±0.2)** | 24[15], 27[14], **21.94(±0.1)** | **288(±18)**, 270 | **13.67(±0.12)** | **18.14(±0.09)** | 150 | - | - |

**This work combing experiments and theory**, *theory only (MoTe₂)*.

Table TS2 – Gate dependence of energy shift of charged Rydberg excitons of all TMDCs.

| Material | A1s+ | A1s- | A2s+ | A2s- | A3s+ | A3s- |
|---|---|---|---|---|---|---|
| MoS$_2$ | - | - | - | - | - | - |
| MoSe$_2$ | −0.75 ± 0.05[7], −1.91 ± 0.06[7], 0.9[6], 0.4[16] | 1.10 ± 0.07[7], 2.9 ± 0.1[7], 1.2[16] | −7.9 ± 0.6[7], −2.9 ± 1.6[7], 3.9[6], 1.8[16] | 11.1 ± 0.6[7], 7.0 ± 1.0[7], 4[16] | −23.8 ± 1.6[7] | 13.7 ± 2.3[7] |
| WS$_2$ | - | - | - | - | - | - |
| WSe$_2$[7] | −0.18 ± 0.02, −1.74 ± 0.03 | 1.2 ± 0.1, 2.3 ± 0.1 | −5.5 ± 0.4, −1.8 ± 0.1 | 4.7 ± 0.3 | - | - |
| MoTe$_2$ | **-0.40±0.10** | **0.28±0.05** | **-8.59±0.54** | **4.23±0.51** | - | - |

Noted quantity is $\frac{d\Delta E}{dE_F}$, where $E_F = \frac{n\pi\hbar^2}{m^*}$ ($n$ is the sheet charge density and $m^*$ is the effective mass).

**This work ($m_e^* = 0.647 m_0, m_h^* = 0.805 m_0$). Effective mass obtained from GW+BSE calculations ($MoTe_2$).**

### O. Computational details.

We first perform density functional theory (DFT) calculations with the Perdew-Burke-Ernzerhof (PBE) generalized gradient approximation[17] with the Quantum Espresso package[18,19], which uses a plane-wave basis set, and norm-conserving pseudopotentials[20,21]. We use a cut-off energy of 125 Ry and include a vacuum of along the out of plane direction, to avoid spurious interactions with repeated unit cells. Both in-plane lattice parameters and atomic coordinates are optimized within the unit cell. A uniform 24x24x1 k-grid is used in the self consistent density calculation, whereas the wave function is generated on a 12x12x1 k-grid. Spin orbit coupling (SOC) is considered. The GW calculations[22–24] are performed with the Berkeley GW code[25,26] using a generalized plasmon pole (GPP) model[24] for the undoped system. Calculations are done with a dielectric cut-off energy of 35 Ry, on a 12x12x1 q-grid with the nonuniform neck subsampling (NNS) scheme[27], with 6,000 states in the summation over unoccupied states, and using a truncated Coulomb interaction[28]. The BSE matrix elements are computed on a uniform 24x24x1 coarse k-grid, then interpolated onto a 288x288x1 fine k-grid, with two valence and four conduction bands. For the doped systems, we use a newly developed plasmon pole model accounting for both dynamical screening effects and local fields effects associated with the free carriers, which is used at both the GW and BSE levels. Further details can be found in Ref. [Champagne et al., submitted]. The doping is introduced as a shift of the Fermi energy above the bottom of the conduction band, corresponding to doping densities ranging from 0 up to $8.7 \times 10^{12} cm^{-2}$. The dielectric matrix, kernel matrix, and BSE Hamiltonian are built for a 48x48x1 k-grid. SOC is not considered in the calculations for the doped systems. The computed intrinsic radiative linewidth is obtained from a Fermi's golden rule following Ref.[29,30].

## P. Computation of exciton dispersion.

We construct BSE Hamiltonian for a series of center-of-mass wavevectors Q in a dense 90 × 90 k-grid. Solving for the lowest energy exciton at each Q yields the exciton dispersion with the effective mass $M_X = 0.87\ m_0$, where $m_0$ is the electron mass.

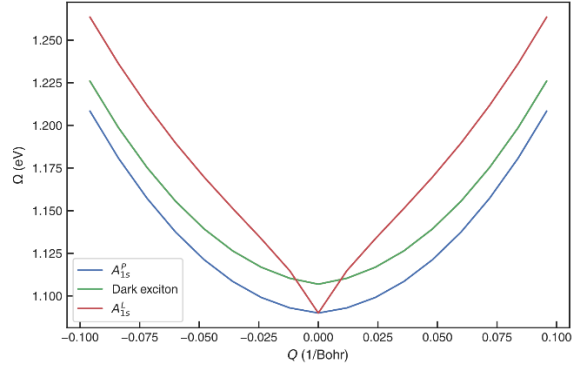

Figure A15. Exciton dispersion for a few lowest energy excitonic states in monolayer MoTe$_2$.

## Q. Computed exciton absorption spectrum as a function of doping density.

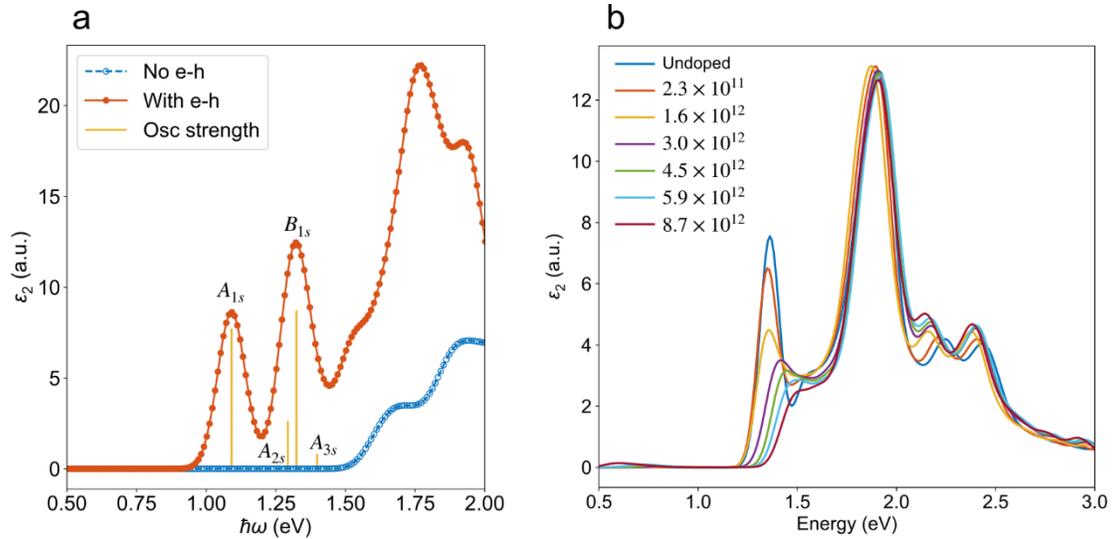

Figure A16. (a) Computed imaginary part of the dielectric function ($\epsilon_2(\hbar\omega)$) for monolayer MoTe2 with and without electron-hole interactions and projected oscillator strength of the different Rydberg excitons (A1s, A2s, B1s, A3s from left to right). (b) Evolution of the imaginary part of the dielectric function as a function of doping density.

## R. Computed exciton wave-function for different doping densities.

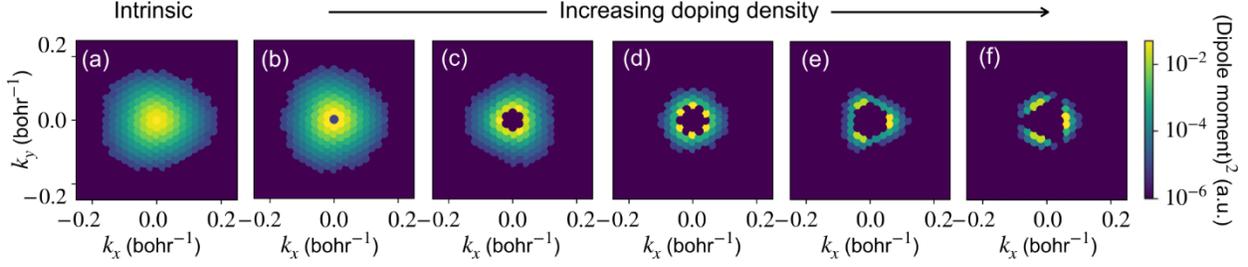

Figure A17. (a)-(f) Evolution of the exciton (A1s) wavefunction as a function of doping density showing Pauli blocking. (b) 2.3 x$10^{11}$ cm$^{-2}$ (c) 1.6 x$10^{12}$ cm$^{-2}$ (d) 3.0 x$10^{12}$ cm$^{-2}$ (e) 4.5 x$10^{12}$ cm$^{-2}$ (f) 5.9x$10^{12}$ cm$^{-2}$.

## S. Computation of trion binding energy.

To obtain charged excitations, we obtain excited-state properties associated with N+3-particle excitations, which consist of one neutral electron-hole pair plus an additional carrier (which we restrict here to be an extra electron). We note that such a description for a trion is applicable for a vanishingly small Fermi surface; for larger carrier doping, one needs to explicitly include the hybridization with of the neutral electron-hole pair with intraband plasmons in the degenerate Fermi see.

We solve for trion excitations by writing a Dyson's equation associated with correlated 3-particle excitations, $L = LKL_0$ where $L_0$ is non-interacting Green's function, $L$ is the interacting Green's function, and $K$ is the interaction kernel. We obtain an equation of motion equivalent to that derived in a prior work[31], and which we will detail in a subsequent manuscript.

We expand our trion wave function in the electron and hole basis in the same way we constructed the exciton wave function

$$|T(n,\boldsymbol{q})\rangle = \sum_{\substack{vc_1c_2 \\ k_1k_2}} B^{n,\boldsymbol{q}}_{\substack{vc_1c_2 \\ k_1k_2}} \hat{c}_{v\boldsymbol{k}_1+\boldsymbol{k}_2-\boldsymbol{q}} \hat{c}^\dagger_{c_2\boldsymbol{k}_2} \hat{c}^\dagger_{c_1\boldsymbol{k}_1} |0\rangle$$

where T is a trion wavefunction with principal quantum number n and wavevector q, B are expansion coefficients, v and c label valence and conduction bands, respectively, k is a wavevector, cˆ is a fermionic destruction operator, and $|0\rangle$ is the many-body ground state. and we arrive at the Dyson's-like equation

$$(E_{c_1\boldsymbol{k}_1} + E_{c_2\boldsymbol{k}_2} - E_{v\boldsymbol{k}_1+\boldsymbol{k}_2-\boldsymbol{q}})B^{n,\boldsymbol{q}}_{\substack{vc_1c_2 \\ k_1k_2}} - \sum_{n'}\langle T(n',\boldsymbol{q})|\widehat{K}|T(n,\boldsymbol{q})\rangle = \Omega^T_{n,\boldsymbol{q}} B^{n,\boldsymbol{q}}_{\substack{vc_1c_2 \\ k_1k_2}}$$

where $\Omega^T_{n,q}$ is the energy of trion at state n and momentum **q** and $E_{mk}$ is the quasiparticle energy for an electron in band m and wavevector **k**. From here, we found the trion binding energy $\Delta E = \Omega^{min}_{exciton} - \Omega^{min}_{trion}$ of 20.6 meV for the A$_{1s}$ state. Our optical spectrum associated with the absorption of excitons and trions is shown in Fig. A18.

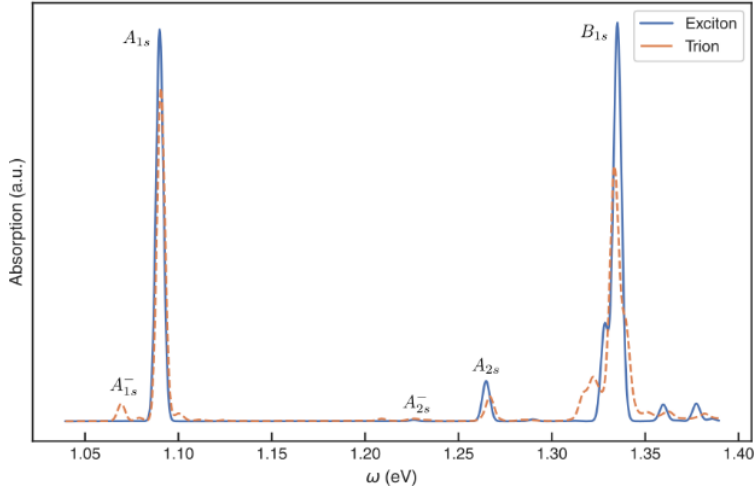

Figure A18. Computed imaginary part of the dielectric function for excitons and trions.

T.  Discussion on importance of MoTe$_2$ optical properties and its Rydberg series.

Well studied TMDCs like Mo and W based sulfides and selenides have band-gaps and optical transitions in the visible spectrum (~600-760 nm). Rydberg excitons associated with these materials are thus at even higher energies. From a technology point of view wavelengths near the silicon band edge (~1.1 eV or ~1100 nm) and telecom band (~1550nm) are very important for the development of silicon-based opto-electronics. With the advent of hybrid platforms where new materials are being integrated into silicon photonics, it is important to find and expand the library of materials which have strong photo-response in these wavelengths. MoTe$_2$ is one of the few 2D materials which in its semiconducting 2H polytype exhibits a ground state excitonic optical transition in the silicon band-edge window. It has thus attracted integration into waveguide-based photonic and opto-electronic applications; for example – light emitting diode[32], waveguide-integrated high-speed[33] and strain-engineered[34] photodetector. It is evident that MoTe$_2$ has attracted significant interest, yet these studies are done for bilayer and bulk samples limiting the achievable performance because the excitonic photo-response should be maximal for a monolayer[35]. A careful and detailed understanding of the photo physics of monolayer MoTe$_2$ at low temperatures under electrostatic doping conditions is lacking which is the primary motivation of selecting this material.

Rydberg excitons in MoTe$_2$ are in a unique part of the electromagnetic spectrum (~800-1000 nm) which is exciting for a lot of applications in the near infrared such as quantum optics with rare-earth ions[36], opto-electronics for health sensing[37] and laser technology[38], including high-power applications[39]. Furthermore, Rydberg excitons provide a way to conveniently study long-range dipole-dipole interactions because of their large size. As the quantum number increases for the Rydberg exciton their average radius $\langle r_n \rangle$ increases (for a hydrogenic system) as per the formula - $\langle r_n \rangle = \frac{1}{2} a_B (3n^2 - l(l+1))$, where $a_B$ is the Bohr radius, $n$ is the principal quantum number and $l$ is the angular momentum. Such an increasing size leads to huge interaction effects and can provide insights into atomic and molecular physics at the single-particle/quantum level. They can also provide excellent sensing capabilities since their wavefunction is exceptionally large and are extra sensitive to the environment[40–42], as well as to their self. Furthermore, Rydberg excitons can be engineered to form an ordered array using Rydberg blockade which is very appealing for quantum simulations. They can also play a vital role in non-linear optics when incorporated into optical cavities to form exciton-polariton modes (due to their large exciton wavefunction and strong dipole-dipole repulsion as compared to 1s excitons)[43,44].

U. **Atomic force microscope image of MoTe$_2$ device.**

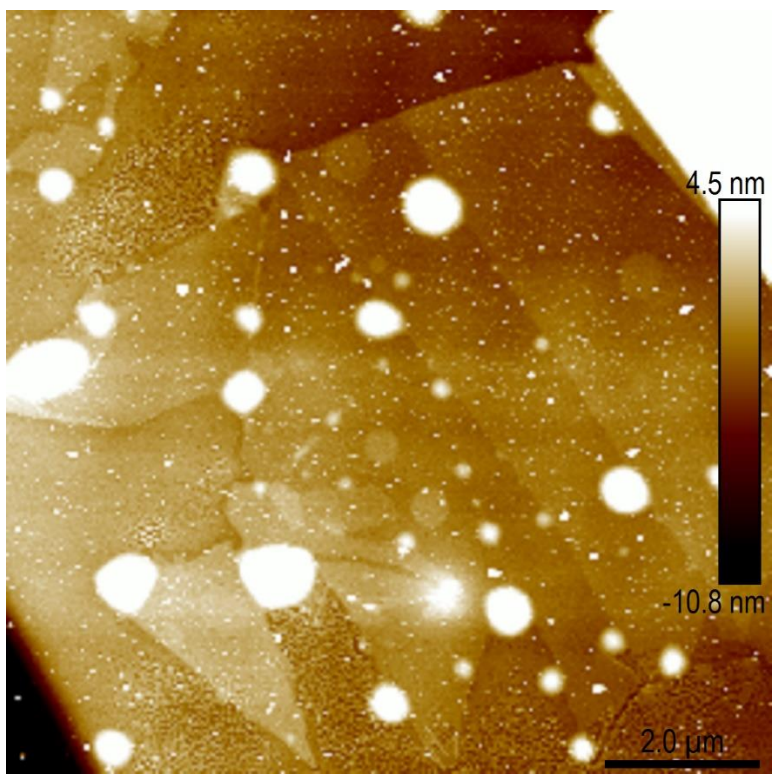

Figure A19. Atomic force microscope image (height sensor) of MoTe$_2$ device for which data was presented in the main manuscript.

**Disclaimer:** Certain commercial equipment, instruments, or materials are identified in this paper in order to specify the experimental procedure adequately. Such identification is not intended to imply recommendation or endorsement by the National Institute of Standards and Technology, nor is it intended to imply that the materials or equipment identified are necessarily the best available for the purpose.

References (SI):

(1) Hill, H. M.; Rigosi, A. F.; Roquelet, C.; Chernikov, A.; Berkelbach, T. C.; Reichman, D. R.; Hybertsen, M. S.; Brus, L. E.; Heinz, T. F. Observation of Excitonic Rydberg States in Monolayer MoS2 and WS2 by Photoluminescence Excitation Spectroscopy. *Nano Lett* **2015**, *15* (5), 2992–2997.

(2) Vaquero, D.; Clericò, V.; Salvador-Sánchez, J.; Martín-Ramos, A.; Díaz, E.; Domínguez-Adame, F.; Meziani, Y. M.; Diez, E.; Quereda, J. Excitons, Trions and Rydberg States in Monolayer MoS2 Revealed by Low-Temperature Photocurrent Spectroscopy. *Communications Physics 2020 3:1* **2020**, *3* (1), 1–8.

(3) Goryca, M.; Li, J.; Stier, A. v.; Taniguchi, T.; Watanabe, K.; Courtade, E.; Shree, S.; Robert, C.; Urbaszek, B.; Marie, X.; Crooker, S. A. Revealing Exciton Masses and Dielectric Properties of Monolayer Semiconductors with High Magnetic Fields. *Nature Communications 2019 10:1* **2019**, *10* (1), 1–12.

(4) Robert, C.; Semina, M. A.; Cadiz, F.; Manca, M.; Courtade, E.; Taniguchi, T.; Watanabe, K.; Cai, H.; Tongay, S.; Lassagne, B.; Renucci, P.; Amand, T.; Marie, X.; Glazov, M. M.; Urbaszek, B. Optical Spectroscopy of Excited Exciton States in MoS2 Monolayers in van Der Waals Heterostructures. *Phys Rev Mater* **2018**, *2* (1), 011001.

(5) Mak, K. F.; He, K.; Lee, C.; Lee, G. H.; Hone, J.; Heinz, T. F.; Shan, J. Tightly Bound Trions in Monolayer MoS2. *Nature Materials 2012 12:3* **2012**, *12* (3), 207–211.

(6) Goldstein, T.; Wu, Y. C.; Chen, S. Y.; Taniguchi, T.; Watanabe, K.; Varga, K.; Yan, J. Ground and Excited State Exciton Polarons in Monolayer MoSe2. *J Chem Phys* **2020**, *153* (7), 071101.

(7) Liu, E.; van Baren, J.; Lu, Z.; Taniguchi, T.; Watanabe, K.; Smirnov, D.; Chang, Y. C.; Lui, C. H. Exciton-Polaron Rydberg States in Monolayer MoSe2 and WSe2. *Nature Communications 2021 12:1* **2021**, *12* (1), 1–8.

(8) Liu, E.; van Baren, J.; Taniguchi, T.; Watanabe, K.; Chang, Y. C.; Lui, C. H. Magnetophotoluminescence of Exciton Rydberg States in Monolayer WSE2. *Phys Rev B* **2019**, *99* (20), 205420.

(9) Chen, S. Y.; Lu, Z.; Goldstein, T.; Tong, J.; Chaves, A.; Kunstmann, J.; Cavalcante, L. S. R.; Woźniak, T.; Seifert, G.; Reichman, D. R.; Taniguchi, T.; Watanabe, K.; Smirnov, D.; Yan, J. Luminescent Emission of Excited Rydberg Excitons from Monolayer WSe2. *Nano Lett* **2019**, *19* (4), 2464–2471.

(10) Wang, T.; Li, Z.; Li, Y.; Lu, Z.; Miao, S.; Lian, Z.; Meng, Y.; Blei, M.; Taniguchi, T.; Watanabe, K.; Tongay, S.; Smirnov, D.; Zhang, C.; Shi, S. F. Giant Valley-Polarized Rydberg Excitons in Monolayer WSe2 Revealed by Magneto-Photocurrent Spectroscopy. *Nano Lett* **2020**, *20* (10), 7635–7641.

(11) He, K.; Kumar, N.; Zhao, L.; Wang, Z.; Mak, K. F.; Zhao, H.; Shan, J. Tightly Bound Excitons in Monolayer WSe2. *Phys Rev Lett* **2014**, *113* (2), 026803.

(12) Wagner, K.; Wietek, E.; Ziegler, J. D.; Semina, M. A.; Taniguchi, T.; Watanabe, K.; Zipfel, J.; Glazov, M. M.; Chernikov, A. Autoionization and Dressing of Excited Excitons by Free Carriers in Monolayer WSe2. *Phys Rev Lett* **2020**, *125* (26), 267401.

(13) Han, B.; Robert, C.; Courtade, E.; Manca, M.; Shree, S.; Amand, T.; Renucci, P.; Taniguchi, T.; Watanabe, K.; Marie, X.; Golub, L. E.; Glazov, M. M.; Urbaszek, B. Exciton States in Monolayer MoSe2 and MoTe2 Probed by Upconversion Spectroscopy. *Phys Rev X* **2018**, *8* (3), 031073.

(14) Yang, J.; Lü, T.; Myint, Y. W.; Pei, J.; Macdonald, D.; Zheng, J. C.; Lu, Y. Robust Excitons and Trions in Monolayer MoTe2. *ACS Nano* **2015**, *9* (6), 6603–6609.

(15) Arora, A.; Schmidt, R.; Schneider, R.; Molas, M. R.; Breslavetz, I.; Potemski, M.; Bratschitsch, R. Valley Zeeman Splitting and Valley Polarization of Neutral and Charged Excitons in Monolayer MoTe2 at High Magnetic Fields. *Nano Lett* **2016**, *16* (6), 3624–3629.

(16) Xiao, K.; Yan, T.; Liu, Q.; Yang, S.; Kan, C.; Duan, R.; Liu, Z.; Cui, X. Many-Body Effect on Optical Properties of Monolayer Molybdenum Diselenide. *Journal of Physical Chemistry Letters* **2021**, *12* (10), 2555–2561.

(17) Perdew, J. P.; Burke, K.; Ernzerhof, M. Generalized Gradient Approximation Made Simple. *Phys Rev Lett* **1996**, *77* (18), 3865.

(18) Giannozzi, P.; Andreussi, O.; Brumme, T.; Bunau, O.; Buongiorno Nardelli, M.; Calandra, M.; Car, R.; Cavazzoni, C.; Ceresoli, D.; Cococcioni, M.; Colonna, N.; Carnimeo, I.; Dal Corso, A.; de Gironcoli, S.; Delugas, P.; Distasio, R. A.; Ferretti, A.; Floris, A.; Fratesi, G.; Fugallo, G.; Gebauer, R.; Gerstmann, U.; Giustino, F.; Gorni, T.; Jia, J.; Kawamura, M.; Ko, H. Y.; Kokalj, A.; Kücükbenli, E.; Lazzeri, M.; Marsili, M.; Marzari, N.; Mauri, F.; Nguyen, N. L.; Nguyen, H. v.; Otero-De-La-Roza, A.; Paulatto, L.; Poncé, S.; Rocca, D.; Sabatini, R.; Santra, B.; Schlipf, M.; Seitsonen, A. P.; Smogunov, A.; Timrov, I.; Thonhauser, T.; Umari, P.; Vast, N.; Wu, X.; Baroni, S. Advanced Capabilities for Materials Modelling with Quantum ESPRESSO. *Journal of Physics: Condensed Matter* **2017**, *29* (46), 465901.

(19) Giannozzi, P.; Baroni, S.; Bonini, N.; Calandra, M.; Car, R.; Cavazzoni, C.; Ceresoli, D.; Chiarotti, G. L.; Cococcioni, M.; Dabo, I.; Dal Corso, A.; de Gironcoli, S.; Fabris, S.; Fratesi, G.; Gebauer, R.; Gerstmann, U.; Gougoussis, C.; Kokalj, A.; Lazzeri, M.; Martin-Samos, L.; Marzari, N.; Mauri, F.; Mazzarello, R.; Paolini, S.; Pasquarello, A.; Paulatto, L.; Sbraccia, C.; Scandolo, S.; Sclauzero, G.; Seitsonen, A. P.; Smogunov, A.; Umari, P.; Wentzcovitch, R. M. QUANTUM ESPRESSO: A Modular and Open-Source Software Project for Quantum of Materials. *Journal of Physics: Condensed Matter* **2009**, *21* (39), 395502.

(20) Hamann, D. R. Optimized Norm-Conserving Vanderbilt Pseudopotentials. *Phys Rev B Condens Matter Mater Phys* **2013**, *88* (8), 085117.

(21) Schlipf, M.; Gygi, F. Optimization Algorithm for the Generation of ONCV Pseudopotentials. *Comput Phys Commun* **2015**, *196*, 36–44.

(22) Hedin, L. New Method for Calculating the One-Particle Green's Function with Application to the Electron-Gas Problem. *Physical Review* **1965**, *139* (3A), A796.

(23) Strinati, G.; Mattausch, H. J.; Hanke, W. Dynamical Aspects of Correlation Corrections in a Covalent Crystal. *Phys Rev B* **1982**, *25* (4), 2867.

(24) Hybertsen, M. S.; Louie, S. G. Electron Correlation in Semiconductors and Insulators: Band Gaps and Quasiparticle Energies. *Phys Rev B* **1986**, *34* (8), 5390.

(25) Deslippe, J.; Samsonidze, G.; Jain, M.; Cohen, M. L.; Louie, S. G. Coulomb-Hole Summations and Energies for GW Calculations with Limited Number of Empty Orbitals: A Modified Static Remainder Approach. *Phys Rev B Condens Matter Mater Phys* **2013**, *87* (16), 165124.

(26) Deslippe, J.; Samsonidze, G.; Strubbe, D. A.; Jain, M.; Cohen, M. L.; Louie, S. G. BerkeleyGW: A Massively Parallel Computer Package for the Calculation of the Quasiparticle and Optical Properties of Materials and Nanostructures. *Comput Phys Commun* **2012**, *183* (6), 1269–1289.

(27) da Jornada, F. H.; Qiu, D. Y.; Louie, S. G. Nonuniform Sampling Schemes of the Brillouin Zone for Many-Electron Perturbation-Theory Calculations in Reduced Dimensionality. *Phys Rev B* **2017**, *95* (3), 035109.

(28) Ismail-Beigi, S. Truncation of Periodic Image Interactions for Confined Systems. *Phys Rev B Condens Matter Mater Phys* **2006**, *73* (23), 233103.

(29) Palummo, M.; Bernardi, M.; Grossman, J. C. Exciton Radiative Lifetimes in Two-Dimensional Transition Metal Dichalcogenides. *Nano Lett* **2015**, *15* (5), 2794–2800.

(30) Katsch, F.; Knorr, A. Excitonic Theory of Doping-Dependent Optical Response in Atomically Thin Semiconductors. *Phys Rev B* **2022**, *105* (4), 045301.

(31) Deilmann, T.; Drüppel, M.; Rohlfing, M. Three-Particle Correlation from a Many-Body Perspective: Trions in a Carbon Nanotube. *Phys Rev Lett* **2016**, *116* (19), 196804.

(32) Bie, Y. A MoTe2-Based Light-Emitting Diode and Photodetector for Silicon Photonic Integrated Circuits. *Nat. Nanotechnol.* **2017**, *12*, 1124–1129.

(33) Flöry, N.; Ma, P.; Salamin, Y.; Emboras, A.; Taniguchi, T.; Watanabe, K.; Leuthold, J.; Novotny, L. Waveguide-Integrated van Der Waals Heterostructure Photodetector at Telecom Wavelengths with High Speed and High Responsivity. *Nat Nanotechnol* **2020**, *15* (2), 118–124.

(34) Maiti, R.; Patil, C.; Saadi, M. A. S. R.; Xie, T.; Azadani, J. G.; Uluutku, B.; Amin, R.; Briggs, A. F.; Miscuglio, M.; van Thourhout, D.; Solares, S. D.; Low, T.; Agarwal, R.; Bank, S. R.; Sorger, V. J. Strain-Engineered High-Responsivity MoTe2 Photodetector for Silicon Photonic Integrated Circuits. *Nat Photonics* **2020**, *14* (9), 578–584.

(35) Lezama, I. G.; Arora, A.; Ubaldini, A.; Barreteau, C.; Giannini, E.; Potemski, M.; Morpurgo, A. F. Indirect-to-Direct Band Gap Crossover in Few-Layer MoTe2. *Nano Lett* **2015**, *15* (4), 2336–2342.

(36) Kindem, J. M.; Ruskuc, A.; Bartholomew, J. G.; Rochman, J.; Huan, Y. Q.; Faraon, A. Control and Single-Shot Readout of an Ion Embedded in a Nanophotonic Cavity. *Nature 2020 580:7802* **2020**, *580* (7802), 201–204.

(37) Biswas, S.; Shao, Y.; Hachisu, T.; Nguyen-Dang, T.; Visell, Y. Integrated Soft Optoelectronics for Wearable Health Monitoring. *Adv Mater Technol* **2020**, *5* (8), 2000347.

(38) Dalir, H.; Koyama, F. 29 GHz Directly Modulated 980 Nm Vertical-Cavity Surface Emitting Lasers with Bow-Tie Shape Transverse Coupled Cavity. *Appl Phys Lett* **2013**, *103* (9), 091109.

(39) Alford, W. J.; Allerman, A. A.; Raymond, T. D. High Power and Good Beam Quality at 980 Nm from a Vertical External-Cavity Surface-Emitting Laser. *JOSA B, Vol. 19, Issue 4, pp. 663-666* **2002**, *19* (4), 663–666.

(40) Xu, Y.; Liu, S.; Rhodes, D. A.; Watanabe, K.; Taniguchi, T.; Hone, J.; Elser, V.; Mak, K. F.; Shan, J. Correlated Insulating States at Fractional Fillings of Moiré Superlattices. *Nature 2020 587:7833* **2020**, *587* (7833), 214–218.

(41) Xu, Y.; Horn, C.; Zhu, J.; Tang, Y.; Ma, L.; Li, L.; Liu, S.; Watanabe, K.; Taniguchi, T.; Hone, J. C.; Shan, J.; Mak, K. F. Creation of Moiré Bands in a Monolayer Semiconductor by Spatially Periodic Dielectric Screening. *Nature Materials 2021 20:5* **2021**, *20* (5), 645–649.

(42) Popert, A.; Shimazaki, Y.; Kroner, M.; Watanabe, K.; Taniguchi, T.; Imamoğlu, A.; Smoleński, T. Optical Sensing of Fractional Quantum Hall Effect in Graphene. *Nano Lett* **2022**, *22* (18), 7363–7369.

(43) Walther, V.; Johne, R.; Pohl, T. Giant Optical Nonlinearities from Rydberg Excitons in Semiconductor Microcavities. *Nature Communications 2018 9:1* **2018**, *9* (1), 1–6.

(44) Gu, J.; Walther, V.; Waldecker, L.; Rhodes, D.; Raja, A.; Hone, J. C.; Heinz, T. F.; Kéna-Cohen, S.; Pohl, T.; Menon, V. M. Enhanced Nonlinear Interaction of Polaritons via Excitonic Rydberg States in Monolayer WSe2. *Nature Communications 2021 12:1* **2021**, *12* (1), 1–7.